\documentclass[11pt]{article}

\pdfoutput=1

\usepackage[T1]{fontenc}
\usepackage[latin9]{inputenc}
\usepackage[a4paper]{geometry}
\usepackage[active]{srcltx}
\usepackage{amsmath}
\usepackage{amssymb}
\usepackage{esint}
\usepackage{ulem}

\makeatletter

\usepackage{textcomp}

\pdfoutput=1 

\usepackage{jheppub}

\usepackage{etoolbox}
    
    \patchcmd{\maketitle}{\@fpheader}{}{}{}

\newcommand*\xbar[1]{%
  \hbox{%
    \vbox{%
      \hrule height 0.5pt 
      \kern0.3ex
      \hbox{%
        \kern-0.0em
        \ensuremath{#1}%
        \kern-0.0em
      }%
    }%
  }%
}

\usepackage{amsfonts}

\usepackage{bbm}

\newcommand{\be}{\begin{equation}}
\newcommand{\ee}{\end{equation}}
\newcommand{\bea}{\begin{eqnarray}}
\newcommand{\eea}{\end{eqnarray}}

\def\cC{\mathcal{C}}

\def\cG{\mathcal{G}}
\def\cH{\mathcal{H}}

\def\cO{\mathcal{O}}

\def\cW{\mathcal{W}}

\title{\boldmath Asymptotic structure of Carrollian limits of Einstein-Yang-Mills theory in four spacetime dimensions}

\author[a]{Oscar Fuentealba,}
\author[a,b]{Marc Henneaux,}
\author[c]{Patricio Salgado-Rebolledo}
\author[a]{and Jakob Salzer}

\affiliation[a]{Universit\'e Libre de Bruxelles and International Solvay Institutes, ULB-Campus Plaine CP231, B-1050 Brussels, Belgium}
\affiliation[b]{Coll\`ege de France, 11 place Marcelin Berthelot, 75005 Paris, France}
\affiliation[c]{Department of Theoretical Physics, Wroclaw University of Science and Technology, 50-370 Wroclaw, Poland}

\emailAdd{oscar.fuentealba@ulb.be}
\emailAdd{marc.henneaux@ulb.be}
\emailAdd{patricio.salgado-rebolledo@pwr.edu.pl}
\emailAdd{jakob.salzer@ulb.be}

\preprint{ULB-MathPhys-22/05}

\abstract{In this paper, three things are done. First, we study from an algebraic point of view the infinite-dimensional BMS-like extensions of the Carroll algebra relevant to the asymptotic structure of the electric and magnetic Carrollian limits of Einstein gravity.  In the course of this study we exhibit by ``Carroll-Galileo duality'' a new infinite-dimensional BMS-like extension of the Galilean algebra and of its centrally extended Bargmann algebra.  Second, we consider the electric Carrollian limit of the pure Einstein theory and indicate that more flexible boundary conditions than the ones that follow from just taking the limit of the Einsteinian boundary conditions are actually consistent.  These boundary conditions lead to a bigger asymptotic symmetry algebra that involves spatial supertranslations depending on three functions of the angles (instead of one).  Third, we  turn to the Carrollian limit of the coupled Einstein-Yang-Mills system. 
An infinite-dimensional color enhancement of the gauge algebra is found in the electric Carrollian limit of the Yang-Mills field, which allows angle-dependent Yang-Mills transformations at spatial infinity, not available in the Einstein-Yang-Mills case prior to taking the Carrollian electric limit. This enhancement does not occur in the magnetic limit.}

\makeatother

\begin{document}
\maketitle \flushbottom


\newpage

\section{Introduction}
The Carroll algebra $\mathcal{C}$ is one of the contractions of the Poincar\'e algebra, obtained by sending the speed of light $c$ to zero (``ultrarelativistic limit'') \cite{LevyLeblond:1965,Bacry:1968zf}.  Its commutation relations are
\begin{eqnarray} && [M_i,M_j]  = \epsilon_{ijk} M_k, \quad  [M_i, B_j]  = \epsilon_{ijk} B_k, \quad  [B_i, B_j]  =  0, \nonumber \\
&& [M_i, P_j]  = \epsilon_{ijk} P_k, \quad [P_i, B_j] = \delta_{ij} E, \quad [M_i , E] = 0 \quad [E, B_i] = 0\label{eq:Carrollalgebra}
\\
&& [P_i, P_j]  = 0, \quad [P_i, E]  = 0 .\nonumber
\end{eqnarray}
where $P_i$, $M_i$, $E$ and $B_i$ are respectively the generators of spatial translations, spatial rotations, time translations and Carroll boosts.

The Carrollian limit of Einstein gravity arose initially as the ``strong coupling limit'' \cite{Isham:1975ur} or ``zero signature limit'' \cite{Teitelboim:1978wv,Henneaux:1981su} of general relativity.  It is relevant to the description of the generic behaviour of the gravitational field in the vicinity of a spacelike singularity \cite{Belinsky:1970ew,Belinsky:1982pk} and even more so when $p$-forms are included \cite{Damour:2000wm,Damour:2002et} (see the review \cite{Belinski:2017fas}).  The geometry of the Carrollian limit was constructed in \cite{Henneaux:1979vn}. Since then, many applications and properties of Carrollian structures have been studied, in particular in the context of the geometry of null surfaces and the BMS symmetry \cite{Dautcourt:1997hb,Duval:2014uoa,Duval:2014uva} (see the recent articles \cite{Donnay:2022aba,Figueroa-OFarrill:2021sxz} for an updated, comprehensive list of references and \cite{Bergshoeff:2022eog} for a review of non-lorentzian theories).

It was shown recently that there are at least two different Carrollian limits of general relativity, an ``electric'' one and a ``magnetic'' one \cite{Henneaux:2021yzg} (see also \cite{deBoer:2021jej,Figueroa-OFarrill:2022mcy,Hansen:2021fxi} for a different perspective on Carrollian gravity theories). The electric Carrollian action corresponds to the zero signature limit of Einstein gravity, whereas the magnetic limit is equivalent to the Carrollian gravity theory found in \cite{Bergshoeff:2017btm} by gauging the Carroll algebra \cite{toappear}.

In  a very interesting paper \cite{Perez:2021abf}, the asymptotic structures of the electric and magnetic limits of the Einstein theory were determined (see \cite{Perez:2022jpr} for the analysis in the presence of negative cosmological constant).  It was shown that the magnetic and electric limits have distinct asymptotic symmetry groups.  While the asymptotic symmetries in the magnetic case are simply the $c \rightarrow 0$ contractions of those of Einstein's theory with the same number of improper \cite{Benguria:1976in} gauge transformations, there is an effective disappearance in the electric case of all the dynamical improper gauge symmetries (boosts and time (super)translations),  because these become in the limit proper gauge transformations  acting trivially on the physical states. 

The boundary conditions studied in \cite{Perez:2021abf} were naturally taken to be the Carrollian contractions of the various boundary conditions considered at spatial infinity for Einstein's theory (\cite{Regge:1974zd,Henneaux:2018cst,Henneaux:2018hdj,Henneaux:2019yax}).    It turns out that in the Carrollian electric limit, a less strict set of boundary conditions can also be consistently imposed. These are not obtainable through the limiting process and lead to an infinite-dimensional enhancement of the Carroll spatial supertranslations.

One purpose of this paper is to analyse these more flexible boundary conditions for the electric Carrollian limit of Einstein gravity.   We explicitly show that while the boosts and the time supertranslations remain pure gauge, as in   \cite{Perez:2021abf}, new space supertranslations introducing further functions of the angles indeed appear.  These involve three functions of the angles and have no analog in the investigations of \cite{Regge:1974zd,Henneaux:2018cst,Henneaux:2018hdj,Henneaux:2019yax} of the  Einstein case.  

In order to shed light on the algebras that can occur asymptotically, we study prior to the asymptotic considerations the Carroll-like contractions of the BMS$_4$ algebra \cite{Bondi:1962px,Sachs:1962wk} from a purely algebraic viewpoint.  This analysis enables one to exhibit, by ``Carroll-Galileo duality'',  a new infinite-dimensional BMS-like extension of the Galilean algebra and of its centrally extended Bargmann algebra. 
We also study the BMS-like extensions of some ideals of the Carroll algebra relevant to the asymptotic symmetries. 

Another purpose of our paper is to include the coupling to the Yang-Mills theory.  This system is of interest in the asymptotic analysis context since it was shown in \cite{Tanzi:2020fmt,Tanzi:2021prq} that there were obstructions to finding at spatial infinity the angle-dependent color transformations exhibited at null infinity \cite{Strominger:2013lka,Barnich:2013sxa,Banerjee:2021uxe}, except in the abelian case studied in \cite{Henneaux:2018gfi}.   We start by considering the Yang-Mills theory on a flat Carroll background.  We show that while the magnetic limit presents the same difficulties as its Lorentzian counterpart, the electric limit allows again for a greater flexibility.  We present boundary conditions that are compatible with the Carroll symmetry and invariant under an angle-dependent color group. 

We then extend these results to the different Carrollian limits of the coupled Einstein-Yang-Mills system, which is a rather   direct task once the flat Carroll case has been understood.

Our paper is organized as follows.  Sections  {\bf \ref{sec:ideals}} and {\bf \ref{sec:BMS-algebra}} study the algebraic structure of the Carroll algebra and its BMS-like extensions. A BMS-like extension of the Carroll algebra is by definition a semi-direct sum of the homogeneous Carroll algebra and an infinite-dimensional abelian algebra parametrized by functions on the sphere (the ``supertranslations'') that contains the ordinary spacetime translations. This is just the analog of the BMS extension of the Poincar\'e group \cite{Bondi:1962px,Sachs:1962wk}.  We use the 3+1  parametrization of the symmetry introduced in \cite{Troessaert:2017jcm} (see also \cite{Henneaux:2018cst}).   We show that there exist at least three inequivalent BMS-like extensions.  By applying the same methods to the non-relativistic limit, we exhibit similarly a third, and to our knowledge new, BMS-like extension of the Galilean algebra and of its centrally extended version, the Bargmann algebra, in addition to the two extensions already constructed in \cite{Batlle:2017yuz}.  We study  next the asymptotic analysis of the electric Carrollian limit of pure Einstein gravity, after a brief survey of the Minkowskian results (Sections {\bf \ref{sec:Asymptotics}} and  {\bf \ref{sec:ElectricCarrolEinstein}}).  We then turn the electric Carrollian limit of the Yang-Mills theory in flat Carroll space (Section {\bf \ref{sec:ElectricCarrollYM}}) and finally to the different Carrollian limits of the combined Einstein-Yang-Mills system (Section {\bf \ref{sec:ElectricCarrollEinsteinYM}}).  Section
 {\bf \ref{sec:Conclusions}} contains concluding comments.

\section{Ideals of the Carroll algebra} \label{sec:ideals}

Among the Carroll transformations, $P_i$ are $M_i$ are kinematical transformations defined within equal time hypersurfaces, while $E$ and $B_i$ are dynamical transformations involving time evolution.  The kinematical transformations form a subalgebra isomorphic to the algebra of Euclidean displacements $iso(3)$. 

The Carroll algebra is not simple and possesses many ideals. For instance, the dynamical transformations $E$ and $B_i$ form an abelian ideal $\mathcal{D}$.  The quotient of the Carroll algebra by the ideal  $\mathcal{D}$ is isomorphic to $iso(3)$,
\be
\frac{\mathcal{C}}{\mathcal{D}} \simeq iso(3).
\ee

The energy $E$ (time translations) by itself also generates a one-dimensional  (abelian) ideal $\mathcal{I}$. The quotient of the Carroll algebra by the ideal  $\mathcal{I}$ is isomorphic to the semi-direct sum of $iso(3)$ and a three-dimensional abelian algebra $t_3$ transforming in the vector representation of $iso(3)$,  denoted $d_3$
\be
\frac{\mathcal{C}}{\mathcal{I}} \simeq iso(3)\oplus_\sigma t_3 \equiv d_3
\ee
(of which $t_3$ is an ideal).

The subalgebra of homogeneous Carroll transformations is spanned by the spatial rotations and the boosts.  It is six-dimensional and isomorphic to $so(3) \oplus_\sigma t_3$.  These transformations grow up at infinity linearly with $r$, while the spatial translations and the time translations tend to constants.

\section{Supertranslations}
\label{sec:BMS-algebra}

\subsection{Carroll-BMS ($\mathcal{C}$-BMS) algebra }\label{sec:Carroll-BMS}

Each of the previous quotient algebras, including the Carroll algebra itself, admits infinite-dimensional extensions by supertranslations.  We now describe extensions for each of them, even though only the extensions of $\mathcal{C}$ and $iso(3)$ are realized through the asymptotic conditions displayed below.

Before turning to these infinite-dimensional extensions, it is helpful to briefly review the Lorentzian case. As is well-known, both at null \cite{Bondi:1962px,Sachs:1962wk} and spatial infinity \cite{Henneaux:2018cst,Henneaux:2018hdj}, the Poincar\'e algebra is extended by supertranslations that are labelled by an arbitrary function on the 2-sphere. These transform as an infinite-dimensional representation of the homogeneous part of the Poincar\'e algebra, i.e., the Lorentz algebra. In a Hamiltonian description around spatial infinity \cite{Henneaux:2018cst,Henneaux:2018hdj}, one naturally finds a parametrization in terms of one even $T(\theta, \varphi)$ and one odd function $W(\theta, \varphi)$ on the 2-sphere that transform as
\begin{equation}
  \label{eq:STtrafo}
	\hat T  = Y^A\partial_A T - 3 b W - \partial_A b \xbar D^A W - b \xbar D_A\xbar D^A W,  \qquad \hat W  = Y^A \partial_A W-bT .
      \end{equation}
In these expressions, $b= b_i n^i$
 parametrizes the boosts and $Y^A$ the rotations, where  the $n_i$'s are the components of the unit normal vector to the 2-spheres -- we refer the reader to the original works for our notation.
   An analysis in hyperbolic coordinates in the neighbourhood of spatial infinity \cite{Troessaert:2017jcm,Henneaux:2018cst} allows one to match this Hamiltonian description involving an even and an odd function on the 2-sphere to the description in terms of the single, unrestricted function on the 2-sphere that one finds at null infinity. 

Let us now turn to the Carroll algebra $\mathcal C$. In a similar way to the Poincar\'e algebra, the Carroll algebra can be extended by Carroll supertranslations, which form an infinite-dimensional representation of the homogeneous Carroll subalgebra and commute among themselves. One expedient manner to obtain the extension is to take the $c \rightarrow 0$ contraction of the BMS$_4$ algebra in a way compatible with the Carrollian contraction of the Poincar\'e algebra.  

In a formulation adapted to the Hamiltonian description, Carroll supertranslations are parametrized by one even function $T(\theta, \varphi)$ and one odd function $W(\theta, \varphi)$ on the $2$-sphere.  The transformation properties of the parameters of supertranslations under boosts can be obtained from \eqref{eq:STtrafo} by the scaling limit $T\rightarrow c T, b\rightarrow cb, W\rightarrow W$, yielding the transformation law
\begin{equation}
		\hat T  = Y^A\partial_A T - 3 b W - \partial_A b \xbar D^A W - b
	\xbar D_A\xbar D^A W , \qquad
	\hat W  = Y^A \partial_A W  . \label{eq:LorentzTransTW}
        \end{equation}
Here, $\xbar D_A$ stands for the covariant derivative associated to the metric $\xbar g_{AB}$ on the 2-sphere.  This contraction reproduces in particular the commutation relations  (\ref{eq:Carrollalgebra}) of the energy and the linear momentum with the Carroll boosts, i.e., $[P_i, B_j] = \delta_{ij} E$ and  $ [E, B_i] = 0$.   

The symmetry algebra defined by (\ref{eq:LorentzTransTW}), together with the commutators  $[M_i,M_j]  = \epsilon_{ijk} M_k$, $[M_i, B_j]  = \epsilon_{ijk} B_k$ and  $[B_i, B_j]  =  0$ of the homogeneous Carroll generators, is denoted by $\mathcal{C}$-BMS.  It arises in the magnetic Carrollian limit of Einstein gravity \cite{Perez:2021abf}.  The charge associated with the $\mathcal{C}$-BMS transformations were found to take the form
\begin{equation}
  Q^M_{\xi}=b_iB^i+\frac{1}{2}b_{ij}M^{ij}+\oint d^2x\sqrt{\xbar g}\,T\,\mathcal T+\oint d^2xW\,\mathcal W\,, 
\end{equation}
($M^{ij} = \epsilon^{ijk} M_k$) where the Poisson-Dirac brackets of the Carroll boosts and the spatial rotations with the Carroll supertranslations are given by 
\begin{eqnarray}
\{B_i,\mathcal{T}(\theta,\phi)\}&=&0\,, \label{CBMS1} \\
\{B_i,\mathcal{W}(\theta,\phi)\}&=&-3n_i \mathcal{T}(\theta,\phi)-\xbar D_A n_i \xbar D^A\mathcal{T}(\theta,\phi)-n_i \xbar D_A\xbar D^A \mathcal{T}(\theta,\phi)\,, \label{CBMS2}\\
\{M^{ij},\mathcal T (\theta,\phi)\}&=&-\xbar D_B(x^{[i}e^{j]B} \mathcal T(\theta,\phi))\,, \label{CBMS3}\\
\{M^{ij},\mathcal W (\theta,\phi)\}&=&-\xbar D_B(x^{[i}e^{j]B} \mathcal W(\theta,\phi))\,, \label{CBMS4}
\end{eqnarray}
in agreement with (\ref{eq:LorentzTransTW}).   

Interestingly, the transformation laws
        \begin{equation}
          \label{eq:opptraf}
\hat{T}=Y^A\partial_A T, \qquad \hat{W}=Y^A\partial_AW-bT
        \end{equation}
        arising from the opposite scaling limit $T\rightarrow T, b\rightarrow \frac{1}{c}b$, $W\rightarrow \frac{1}{c}W$ with $c \rightarrow \infty$ define one BMS extension of the Galilean algebra (reproducing in particular the Galilean relation $[P_i, B_j] = 0$ and  $ [E, B_i] = P_i$).  The comparison of this BMS extension with the extensions constructed in \cite{Batlle:2017yuz} is discussed in the next subsection.

\subsection{$d_3$-BMS  algebra }

The ideals of the Carroll algebra $\mathcal C$ listed above can be extented to include supertranslations.   One can for instance enlarge the ideal generated by the time translation generator $E$ to include all time supertranslations.   

If one takes the quotient of the algebra $\mathcal{C}$-BMS by this ideal, one gets an extension of $d_3$ by space supertranslations, i.e., the algebra generated by $B_i$, $M_{ij}$ and $\mathcal W$ with (\ref{CBMS2}) replaced by
\be
\{B_i,\mathcal{W}(\theta,\phi)\} = 0.
\ee
We call this algebra the $d_3$-BMS  algebra.  It is not realized through the asymptotic conditions given below.

\subsection{$iso(3)$-BMS algebra } \label{sec:iso3-BMS}
Similarly, one can enlarge the ideal generated by the time translation generator $E$ and the boosts $B_i$ to include all time supertranslations. This time, the quotient of the Carroll-BMS algebra by this ideal is spanned by $M_{ij}$ and the spatial supertranslations $\mathcal W$, with the above commutation relations.

This algebra is called the $iso(3)$-BMS algebra and arises in the electric Carrollian limit of Einstein gravity \cite{Perez:2021abf}.

\subsection{Extended $iso(3)$-BMS algebra } \label{sec:iso3-BMS-Extended}

Another infinite-dimensional extension of $iso(3)$ relevant to the asymptotic analysis of the electric limit of Einstein theory is obtained by adding more spatial supertranslations, which we denote $\mathcal W_A(\theta, \phi)$ (parametrized by $I^A$)  and which commute with the spatial supertranslations parametrized by $W$. While the $W$'s are odd functions on the sphere, the $ I^A$ are even functions. 

The transformation of the functions $I^A$ under rotations is
\be
\hat I^{A}=Y^B\partial_B I^{A}-I^{B}\partial_BY^A \,,
\ee
leading to the corresponding brackets for the generators $\cW_A$,
\begin{align}
\{M^{ij},\cW_A (\theta,\phi)\}&=\xbar D_B(x^{[i}e^{j]B} \cW_A(\theta,\phi))-x^{[i}\xbar D_Ae^{j]B}\cW_B (\theta,\phi)\,.
\end{align}

This algebra is called the extended $iso(3)$-BMS algebra. It cannot be obtained by contraction of the BMS$_4$ algebra since it contains more spatial supertranslations.  These are parametrized by one odd function ($W$) and two even functions ($I^A$) on the sphere.

\subsection{Spherical harmonics presentation}

\subsubsection{BMS$_4$ algebra}

The supertranslation generators $\mathcal T$ and $\mathcal W$ can be decomposed in terms of spherical harmonics,
\be \mathcal T = \sum_{l \geq 0, \hbox{\scriptsize{even}}} \; \sum _{m=-l}^{m=l} P_{lm} Y^l_m \, , \qquad \mathcal W = \sum_{l \geq 1, \hbox{\scriptsize{odd}}} \; \sum _{m=-l}^{m=l} P_{lm} Y^l_m \, ,
\ee
The BMS$_4$ commutation rules take then the form \cite{Sachs:1962zza}
\be
[B_i, P_{lm}] = \sum_{l'} \sum_{m'} \left(C_i\right)_{lm}^{l'm'} P_{l'm'}  \label{eq:CommBP}
\ee
in addition to the commutation relations of the homogeneous Lorentz group $[M_i, M_j]  = \epsilon_{ijk} M_k$, 
$ [M_i, B_j]  = \epsilon_{ijk} B_k$, 
$ [B_i, B_j]  = - \epsilon_{ijk} M_k$  and the commutation relations $[M_{i}, P_{lm}]$
that express that the spherical harmonics $\{Y^l_m\}$ ($m = - l, \cdots, l$) form a basis of the spin-$l$ representation of the rotation group, so that $\mathcal T$ and $\mathcal W$ transform as scalar functions on the sphere.

The detailed form of the structure constants $ \left(C_i\right)_{lm}^{l'm'}$ will not be needed here.  It can be found in  \cite{Sachs:1962zza} (after making the appropriate normalization of the $P_{lm}$'s explained in \cite{Troessaert:2017jcm} and \cite{Henneaux:2018cst}).  The only property of the $ \left(C_i\right)_{lm}^{l'm'}$'s that we will need is 
\be
\left(C_i\right)_{lm}^{l'm'} = 0 \hbox{ unless $l = l' \pm 1$},
\ee
so that \eqref{eq:CommBP} involves only two contributions,
\be
[B_i, P_{lm}] =  \sum_{m'} \left(C_i\right)_{lm}^{l-1,m'} P_{l-1,m'}  + \sum_{m'} \left(C_i\right)_{lm}^{l+1,m'} P_{l+1,m'} \label{eq:CommBPBis}
\ee
Furthermore $\vert m' -m \vert $ must be at most equal to one.

Irreducible representations of the Lorentz algebra ($so(3,1)$)  have been systematically investigated in  \cite{Naimark62,Gel'fand63,HarishChandra47}. They are characterized by two numbers.  In the work of \cite{Naimark62,Gel'fand63}, which we follow, these two numbers are denoted $l_0$ and $l_1$, and the corresponding representation is denoted $(l_0, l_1)$.   The first number $ l_0$ is a non-negative integer or half-integer and is the minimum $so(3)$-spin occurring in the decomposition of the representation of the Lorentz algebra according to its $so(3)$-subalgebra.  The second number $l_1$ is an arbitrary complex number.  When $l_1 - l_0$ is a strictly positive integer, the representation is finite-dimensional and $l_1$ is equal to the maximum $so(3)$-spin plus one.  There exist then another representation of $so(3,1)$ characterized by the ``dual'' values $l'_0 = l_1$ and $l'_1 = l_0$.  This representation $(l_1, l_0)$  is called the ``tail'' of the finite-dimensional representation $(l_0,l_1)$.  It is infinite-dimensional since $l'_1 - l'_0 = -(l_1 - l_0)$ is strictly negative.

The two numbers $l_0$ and $l_1$ determine the structure constants $ \left(C_i\right)_{lm}^{l'm'}$  
and the two $so(3,1)$ Casimirs \cite{Naimark62,Gel'fand63},
\be
C_1 \equiv J^2  \equiv   \frac12 M^{\alpha \beta} M_{\alpha \beta} =  {\mathbf  M}^2 - {\mathbf B}^2 = -( l_0^2 + l_1^2) + 1
\ee
and 
\be
C_2 \equiv - \frac12 \epsilon^{\alpha \beta\gamma \delta} M_{\alpha \beta} M_{\gamma \delta} = - {\mathbf M} \cdot {\mathbf B}  = - i l_0 l_1
\ee
where $M_{\alpha \beta}$ are the Lorentz generators in covariant form ($M_{12} = M_3$ etc).
This formula shows that the Casimirs are  invariant under  the exchange of $l_0$ with $l_1$, $(l_0, l_1) \rightarrow (l_1, l_0)$, and thus, cannot distinguish the corresponding  (in general distinct)  irreducible representations.  In particular, it cannot distinguish between a finite-dimensional representation and its tail, which have the same Casimirs.

The representation of the Lorentz group given by  the supertranslations is not irreducible, but is indecomposable \cite{Sachs:1962zza}.  There is a four-dimensional invariant subspace,   characterized by the values $l_0=0$ and $l_1 = 2$.  This is just the standard vector representation, spanned by ordinary spacetime translations with  maximum $so(3)$-spin equal to $1$. The Casimirs are  $C_1 = - 3$, $C_2 = 0$.  The quotient representation of the supertranslations by the translations is infinite-dimensional and isomorphic to the irreducible representation with dual values $l_0=2$ and $l_1 = 0$.  This infinite-dimensional representation is thus the ``tail'' of the finite-dimensional vector representation.  

We close this brief survey of the irreducible representations of the Lorentz algebra by giving explicitly the action of $B_3$ in the irreducible representation $(l_0,l_1)$.  To write this action, it is convenient to decompose the representation in terms of irreducible representations of the compact $so(3)$ subalgebra.  The representations that occur have $so(3)$-spin equal to $l_0$, $l_0 + 1$, $l_0 +2$, etc and this never stops unless $l_1- l_0$ is a positive integer. In a standard spin-basis $\{\xi_{lm}\}$, the action of $B_3$ reads
\begin{eqnarray}
&&iB_3 \xi_{lm} = c_l \sqrt{l^2 - m^2} \xi_{l-1,m} - a_l m \xi_{l, m} - c_{l+1} \sqrt{(l+1)^2 - m^2} \xi_{l+1,m} \label{eq:F1}\\
&& l = l_0, l_0 + 1, \cdots, \qquad m = -l, -l+1 , \cdots, l-1, l,
\end{eqnarray}
where
\be
a_l = \frac{i l_0 l_1}{l (l+1)}, \qquad c_l = \frac{i}{l} \sqrt{\frac{(l^2- l_0^2)(l^2 - l_1^2)}{4 l^2-1}}. \label{eq:EqnsAC}
\ee
The relation (\ref{eq:F1}) involves a specific choice of relative $l$-dependent normalization of the $\{\xi_{lm}\}$, made such that it is the same set of coefficients $c_l$ that characterizes the component of $B_3 \xi_{lm}$ along $\xi_{l-1,m} $and  $\xi_{l+1,m}$.  Note that $a_l$ is equal to zero whenever $l_0$ or $l_1$ vanishes.

One can bring the commutation relations (\ref{eq:CommBP}) with $i=3$, $l=0$ and $l=1$ to the form (\ref{eq:F1}) with $l_0=0$, $l_1 = 2$.  Similarly, modulo terms involving $P_{1m}$ when $l=2$, one can bring the commutation relations (\ref{eq:CommBP}) with $i=3$ and $l\geq 2$ to the form (\ref{eq:F1}) with $l_0=2$, $l_1 = 0$.

\subsubsection{$\mathcal{C}$-BMS algebra}
With the separation of the spherical harmonics into even components  (containing the time translations) and odd components (containing the space translations), the Carroll contraction of the BMS$_4$ algebra is direct.  Indeed, to get from the Poincar\'e algebra to the Carroll algebra, one must rescale differently the time translations ($l=0$) and the space translations ($l=1$), 
\be E  \rightarrow c E, \qquad P_i \rightarrow P_i
\ee
together with $B_i \rightarrow c B_i$ ($c \rightarrow 0$).  This can  consistently be extended to all the supertranslations as follows,
\be P_{lm}  \rightarrow c P_{lm} \quad \hbox{for $l$ even, $l \not=0$} , \qquad  P_{lm}  \rightarrow  P_{lm} \quad \hbox{for $l$ odd}
\ee
leading to the $\mathcal{C}$-BMS algebra of the rescaled generators in the limit $c \rightarrow 0$ (rescaled generators kept fixed),
\be
[\tilde B_i, \tilde P_{lm}] = \sum_{l'} \sum_{m'} \left(C_i\right)_{lm}^{l'm'} \tilde P_{l'm'} \quad  \hbox{for $l$ odd}, \qquad [\tilde B_i, \tilde P_{lm}] = 0 \quad  \hbox{for $l$ even} 
\ee
with $ \left(C_i\right)_{lm}^{l'm'} $ unchanged for $l$ odd, and being equal to zero for $l$ even.  Here 
\be
\tilde B_i = c B_i\, , \qquad \tilde P_{lm} = c P_{lm} \quad \hbox{for $l$ even, $l \not=0$} , \qquad  \tilde P_{lm} =  P_{lm} \quad \hbox{for $l$ odd.}
\ee
From now on, we shall drop the tildes and keep the same symbol for the rescaled generators since no confusion should arise.  So we rewrite the above brackets as
\be
[B_i, P_{lm}] = \sum_{l'} \sum_{m'} \left(C_i\right)_{lm}^{l'm'} P_{l'm'} \quad  \hbox{for $l$ odd}, \qquad [B_i, P_{lm}] = 0 \quad  \hbox{for $l$ even} \label{eq:CommBPCarroll}
\ee
The same convention will be adopted below when we apply other rescalings to get the other Carrollian (and Galilean) contractions.

The splitting of the supertranslations into even and odd parts, natural from the Hamiltonian description, is particularly well adapted to the Carroll contraction.

 The homogeneous Carroll algebra spanned by $M_i,B_i$, being isomorphic to $\mathfrak{iso}(3)$, has the two Casimirs
        \begin{equation}
          \label{eq:CasiCarroll}
          C_1=B_iB_i\qquad C_2=B_iM_i\,.
        \end{equation}
        The supertranslation representation \eqref{eq:LorentzTransTW} ($\Leftrightarrow$ \eqref{eq:CommBPCarroll}) has $C_1=C_2=0$.
        
 One can imagine different rescalings of the supertranslations, where the power of $c$ depends on $l$.  The only restriction is that these should reproduce the rescalings $E  \rightarrow c E$, $P_i \rightarrow P_i$ and yield a well-defined limit consistent with $B_i \rightarrow c B_i$ when $c \rightarrow 0$. 
 One possibility is to take
 \be P_{lm}  \rightarrow c^{1-l} P_{lm}. \label{eq:AltResc}
\ee      
In that case, one gets
\be
[B_i, P_{lm}] =  \sum_{m'} \left(C_i\right)_{lm}^{l-1,m'} P_{l-1,m'} \label{eq:CommBPCarrollBis}
\ee
with unchanged $ \left(C_i\right)_{lm}^{l-1,m'} $ but with $ \left(C_i\right)_{lm}^{l+1,m'} =0$.   This representation of the homogeneous Carroll group has also its Casimirs both equal to zero, but is inequivalent to the previous one.   Both representations are not irreducible, but  indecomposable.  In the case of $\mathcal{C}$-BMS, each subspace with even $\mathfrak{so}(3)$-spin $2k$ ($k$ arbitrary integer) is invariant and form the finite-dimensional spin-$2k$ representation of $\mathfrak{iso}(3)$.  Once one takes the quotient by the adjacent even spins, the subspaces with definite odd $\mathfrak{so}(3)$-spin are also invariant and provide the finite-dimensional representations of $\mathfrak{iso}(3)$ with odd spin.  In the case when the supertranslations transform as in \eqref{eq:CommBPCarrollBis}, one arrives at the same content in terms of irreducible representations, but the quotients are nested, in the sense that the representation with spin $j$ is obtained by taking the quotient of the subspace with $\mathfrak{so}(3)$-spin $j$ by the subspaces with lower $\mathfrak{so}(3)$-spin $j'<j$.

The infinite-dimensional algebras $d_3$-BMS and $iso(3)$-BMS are easy to describe in terms of a spherical harmonic decomposition of the supertranlation generators, restricted in that case to the odd generator $\mathcal W(\theta ,\phi)$, since only the rotation subgroup $so(3)$ acts non trivially.  Similarly, the extended $iso(3)$-BMS has additional supertranslantions transforming as vector fields on the sphere, which are most conveniently expanded in vector spherical harmonics.

\subsubsection{A note on the Galilean contractions}
The Hamiltonian description of the BMS$_4$ algebra is also adapted to  the study of the Galilean contraction \be B_i \rightarrow  \frac{1}{c} B_i, \qquad E  \rightarrow  E, \qquad P_i \rightarrow \frac{1}{c} P_i
\ee
($c \rightarrow \infty$).   Again, there exist various possibilities.  One is the direct analog of the $\mathcal{C}$-BMS algebra and corresponds to  \eqref{eq:opptraf}.  It reads
\be P_{lm}  \rightarrow  P_{lm} \quad \hbox{for $l$ even} , \qquad  P_{lm}  \rightarrow \frac{1}{c} P_{lm} \quad \hbox{for $l$ odd} \label{eq:GalileanScaling0}
\ee
leading to the Galilean-BMS algebra
\be
[B_i, P_{lm}] = \sum_{l'} \sum_{m'} \left(C_i\right)_{lm}^{l'm'} P_{l'm'} \quad  \hbox{for $l$ even}, \qquad [B_i, P_{lm}] = 0 \quad  \hbox{for $l$ odd} \label{eq:CommBPGalileo}
\ee
with $ \left(C_i\right)_{lm}^{l'm'} $ unchanged for $l$ even, and being equal to zero for $l$ odd.

This contracted algebra differs from the algebras $\mathfrak{nrbms}^{\pm}$ considered in \cite{Batlle:2017yuz} and is thus new.  The  algebra  $\mathfrak{nrbms}^{+}$ corresponds to a rescaling analogous to \eqref{eq:AltResc},
 \be P_{lm}  \rightarrow c^{-l} P_{lm}, \label{eq:AltRescGal}
\ee      
leading to 
\be
[B_i, P_{lm}] =  \sum_{m'} \left(C_i\right)_{lm}^{l+1,m'} P_{l+1,m'} \label{eq:CommBPGalileoBis}
\ee
with unchanged $ \left(C_i\right)_{lm}^{l+1,m'} $ but with $ \left(C_i\right)_{lm}^{l-1,m'} =0$. 
As to the  algebra  $\mathfrak{nrbms}^{-}$, it corresponds to the rescaling
 \be P_{00} \rightarrow c^2 P_{00}, \qquad P_{lm}  \rightarrow c^{l} P_{lm} \quad (l \not= 0) \label{eq:AltRescGal}
\ee    
leading to  
\be
[B_i, P_{1m}] =0, \quad [B_i, P_{lm}] =  \sum_{m'} \left(C_i\right)_{lm}^{l-1,m'} P_{l-1,m'} \quad (l \not= 1) \label{eq:CommBPGalileoTer}
\ee
with unchanged $ \left(C_i\right)_{lm}^{l-1,m'} $ ($l \not=1$) but with $ \left(C_i\right)_{lm}^{l+1,m'} =0$. [The Galilean algebra can also be obtained from the Poincar\'e algebra through the rescalings $B_{i} \rightarrow \frac{1}{c} B_i$, $E \rightarrow c^2 E$, $P_i \rightarrow c P_i$.]

In fact, the paper \cite{Batlle:2017yuz}, to which we refer for the details, studied the extensions of the Bargmann algebra, which is the central extension of the Galilean algebra.  They considered scalings equivalent to the ones considered here, adapted to the central extension.
This raises the question as to whether the Galilean-BMS algebra \eqref{eq:CommBPGalileo} admits a central extension that would make it another infinite-dimensional BMS-extension of the Bargmann algebra.  

The answer is affirmative, as can be seen by taking the appropriate limit of the direct sum BMS$_4 \oplus u(1)$ of the BMS$_4$ algebra with the abelian algebra $u(1)$, the generator of which is denoted by $C$.   One thus has prior to contraction
\be
[B_i, C] = 0, \qquad [P_{lm}, C] = 0, \qquad [M_i, C] = 0
\ee
We then ``twist'' the zero mode sector spanned by the generators $(P_{00}, C)$ that commute with $M_i$ through the redefinitions
\be
E = P_{00} +  C , \quad   Z = \frac{1}{2}(P_{00} - C ),   \quad \Leftrightarrow \quad P_{00} = \frac12 E + Z, \quad C = \frac12 E - Z
\ee
Finally, we perform the standard rescalings $B_i \rightarrow \tilde B_i = \frac{1}{c} B_i$, $M_i \rightarrow \tilde M_i = M_i$ and \eqref{eq:GalileanScaling0} for $l>0$, together with the zero-mode rescaling 
\be E \rightarrow \tilde E = E, \qquad Z \rightarrow \tilde Z = \frac{1}{c^2} Z \, .
\ee
This brings the algebra of the rescaled generators with the boosts to the form (dropping the tildes)
\begin{eqnarray}
&& [ B_i,  E] = \sum_{m} \left(C_i\right)_{00}^{1m} P_{1m} , \qquad [B_i, Z]= 0 \\
&& [ B_i,  P_{lm}] = \sum_{l'} \sum_{m'} \left(C_i\right)_{lm}^{l'm'}  P_{l'm'} \: \;   \hbox{for $l$ even $>0$}, \\
&& [ B_i, P_{1m}] =  \left(C_i\right)_{1m}^{00} Z  \\
&&  [ B_i,  P_{lm}] = 0 \; \;  \hbox{for $l$ odd $>1$} 
\label{eq:CommBPGalileoV0}
\end{eqnarray}
and provides indeed a BMS-like extension of the Bargmann algebra different from those of~\cite{Batlle:2017yuz}.

\section{Brief overview of boundary conditions in Einstein gravity} \label{sec:Asymptotics}

After these algebraic preliminaries, we now turn to the asymptotic analysis of Carrollian gravities, starting with a brief overview of Einstein gravity.

The Hamiltonian action of Einstein gravity in four spacetime dimensions reads
\be
S[g_{ij},\pi^{ij},N,N^i]=\int dt \left[\int d^{3}x \left(\pi^{ij}\dot{g}_{ij}-N \mathcal{H}- N^{i} \mathcal{H}_{i}\right)-B_{\infty}\right]\,.
\ee
Here, $N$ and $N^i$ stand for the lapse and shift functions, respectively. The variation of the Hamiltonian action with respect to these functions imposes the following constraints on the momentum $\pi^{ij}$ and the 3-dimensional metric $g_{ij}$
\begin{eqnarray}
\mathcal{H}&=&\frac{1}{\sqrt{g}}\left(\pi^{ij}\pi_{ij}-\frac{1}{2}\pi^2\right)-\sqrt{g}R\approx 0\,,\\
\mathcal{H}_i&=&-2\nabla^j\pi_{ij}\approx 0\,.
\end{eqnarray}
The boundary term at spatial infinity $B_\infty$ depends on the boundary conditions and it turns out to be the standard ADM energy when the lapse and shift functions behave asymptotically as $N\rightarrow 1$ and $N^i\rightarrow 0$, respectively \cite{Regge:1974zd}. The action of an arbitrary diffeomorphism  $\xi^{\mu}=(\xi^{\perp}\equiv\xi,\xi^i)$ on the dynamical fields yields the following infinitesimal transformation laws
\begin{eqnarray}
\delta_{\xi,\xi^i} g_{ij}&=&\frac{2\xi}{\sqrt{g}}\left(\pi_{ij}-\frac{1}{2}g_{ij}\pi\right)+\mathcal{L}_{\xi}g_{ij}\,, \label{eq:TransfEinsteinG}\\
\delta_{\xi,\xi^i} \pi^{ij}&=&-\frac{\xi}{\sqrt{g}}\left(R^{ij}-\frac{1}{2}g^{ij}R\right)+\frac{\xi g^{ij}}{2\sqrt{g}}\left(\pi^{mn}\pi_{mn}-\frac{\pi^2}{2}\right) \nonumber \\
&&-\frac{2\xi}{\sqrt{g}}\left(\pi^{im}\pi^j_m-\frac{1}{2}\pi^{ij} \pi\right)+\sqrt{g}\left(\nabla^i\nabla^j \xi-g^{ij}\nabla_m\nabla^m\xi\right)+\mathcal{L}_{\xi}\pi^{ij}\,,\label{eq:TransfEinsteinPi}
\end{eqnarray}
where $\mathcal{L}_\xi$ denotes the spatial Lie derivative, which acts on the fields as
\begin{eqnarray}
\mathcal{L}_{\xi}g_{ij}&=&\xi^k\partial_kg_{ij}+\partial_{i}\xi^kg_{kj}+\partial_{j}\xi^kg_{ki}\,,\\
\mathcal{L}_{\xi}\pi^{ij}&=&\partial_k (\xi^k \pi^{ij})-\partial_k\xi^i \pi^{jk}-\partial_k\xi^j \pi^{ik}\,.
\end{eqnarray}

There exist different sets of boundary conditions at spatial infinity that have been proposed in the literature \cite{Regge:1974zd,Henneaux:2018cst,Henneaux:2018hdj}.  For all of them, 
the fall-off of the dynamical fields in spherical polar coordinates is given by 
\begin{align}
    g_{rr} &= 1 + \frac {\xbar h_{rr}} {r} +\frac {h^{(2)}_{rr}} {r^2} + \mathcal{O}\left(r^{-3}\right)\,,\\
    g_{rA} &= \xbar{\lambda}_A+\frac{\xbar h_{rA}}{r} + \frac{h^{(2)}_{rA}}{r^2} + \mathcal{O}\left(r^{-3}\right)\,,\\
    g_{AB} &= r^2 \xbar g_{AB} + r  \xbar h_{AB} +h^{(2)}_{AB} + \mathcal{O}\left(r^{-1}\right)\,, \\
    \pi^{rr} &=  \xbar \pi^{rr} +\frac{\pi^{(2)}_{rr}}{r}+ \mathcal{O}\left(r^{-2}\right)\,, \\
    \pi^{rA} &= \frac{\xbar \pi^{rA}}{r} + \frac{\pi^{(2)rA}}{r^2} + \mathcal{O}\left(r^{-3}\right)\,,\\
    \pi^{AB} &=  \frac{\xbar \pi^{AB}}{r^2}+  \frac{\pi^{(2)AB}}{r^3}+ \mathcal{O}\left(r^{-4}\right)\,.
\end{align}
What distinguishes the different sets of boundary conditions are the conditions imposed on the leading orders $\xbar h_{ij}$, $ \xbar \pi^{ij}$ in the expansion of the fields, which must be imposed for the symplectic structure to be finite.  The Carroll contractions of the boundary conditions proposed in \cite{Regge:1974zd,Henneaux:2018cst} have been analysed in detail in \cite{Perez:2021abf}, with the following conclusions: (i) the Carrollian limit of the boundary conditions of \cite{Regge:1974zd} leads to the finite-dimensional asymptotic symmetry algebras $\mathcal C$ in the magnetic case, and $iso(3)$ in the electric one; (ii) the Carrollian limit of the boundary conditions of \cite{Henneaux:2018cst} leads to the infinite-dimensional asymptotic symmetry algebras $\mathcal C$-BMS  in the magnetic case, and $iso(3)$-BMS  in the electric one.  Since  the Carroll contractions of the boundary conditions of \cite{Regge:1974zd,Henneaux:2018cst} are fully understood, we shall focus here on the boundary conditions proposed in \cite{Henneaux:2018hdj}.

\subsection{Boundary conditions of \cite{Henneaux:2018hdj}} \label{subsec:twisted}
The third set of boundary conditions at spatial infinity compatible with finiteness of the symplectic structure that has been proposed in the literature is also BMS invariant, as are the ones of \cite{Henneaux:2018cst}. 
It was introduced in \cite{Henneaux:2018hdj} (see also \cite{Henneaux:2019yax} for more information).  The corresponding boundary conditions differ from the  boundary conditions of \cite{Regge:1974zd} (which do not have the BMS group as asymptotic symmetry group) by an improper gauge transformation and for that reason, are sometimes called diffeomorphism-twisted parity conditions.

In spherical coordinates, the boundary conditions read \cite{Henneaux:2018hdj,Henneaux:2019yax}
\begin{align}
	\xbar h_{rr}&= \text{even}\,,\label{eq:TwistedE1}\\
	\xbar \lambda_A &= (\xbar \lambda_A)^{\text{odd}}+\xbar D_A \zeta_r-\xbar \zeta_A\,,\\ 
    \xbar h_{AB} &= (\xbar h_{AB})^{\text{even}}+\xbar D_A \xbar \zeta_B+ \xbar D_B \xbar \zeta_A+2\xbar g_{AB}\zeta_r\,,\\ 
    \xbar \pi^{rr}&= (\xbar \pi^{rr})^{\text{odd}}-\sqrt{\xbar g}\,\xbar \triangle V\,,\\
    \xbar \pi^{rA}&= (\xbar \pi^{rA})^{\text{even}}-\sqrt{\xbar g}\,\xbar D^A V\,,\\
    \xbar \pi^{AB}&= (\xbar \pi^{AB})^{\text{odd}}+\sqrt{\xbar g}(\xbar D^A \xbar D^B V-\xbar g^{AB}\xbar \triangle V)\,,\label{eq:TwistedE2}
\end{align}
where $\zeta_i dx^i=\zeta_r (\theta, \phi) dr+r\xbar \zeta_A(\theta, \phi)  dx^A$, $\xbar D_A$ is the covariant derivative on the 2-sphere, and $\xbar \triangle$ is the Laplace operator on the 2-sphere. The function $\zeta_r$ is odd, while the angular component $\xbar \zeta_A$ and the function $V$ are even under the antipodal map. 

The condition $\xbar \lambda_A = 0$ (which implies $\xbar \zeta_A=\xbar D_A\zeta_r$) is also imposed to insure integrability of the boost charges, but since it can be relaxed in the Carroll electric limit, we keep this term here.

\subsection{Asymptotic symmetries and BMS$_4$ algebra}
This set of asymptotic conditions is preserved  by the following surface deformation parameters
\begin{eqnarray}\label{eq:diff}
\xi&=&br +T+\text{``more''}+\mathcal{O}\left(r^{-2}\right)\,,\\
\xi^r&=&W+\mathcal{O}\left(r^{-1}\right)\,,\\
\xi^A&=&Y^A+\frac{1}{r}(\xbar D^A W+\text{``more''})+\mathcal{O}\left(r^{-2}\right)\,.
\end{eqnarray}
Here:
\begin{itemize}
\item Lorentz boosts are generated by $b=b_i n^i$ (with $b_i$ arbitrary constants).  Being pure vector spherical harmonic, the function $b$ obeys the equation $\xbar D_A \xbar D_B b+ \xbar g_{AB}b=0$.
\item Spatial rotations are generated by the vectors $Y^A=\frac{1}{2}b_{ij}x^ie^{jA}$ (with $b_{ij}=-b_{ji}$ constant and $e^{jA}$ vectors tangent to the unit sphere), which are Killing vectors of the round 2-sphere, i.e., $\xbar D_A Y_B+\xbar D_B Y_A=0$. 
\item The parameters $T$ and $W$ are respectively arbitrary even and odd functions on the 2-sphere under the antipodal map $\theta \rightarrow \pi-\theta$ and $\phi\rightarrow \phi+\pi$. They generate all the supertranslations.
\item The terms ``more'' correspond to correcting improper diffeomorphisms that preserve the condition $g_{rA}=\mathcal{O}(r^{-1})$, or equivalently $\xbar \lambda_A=0$, which makes the Lorentz boosts canonical transformations. 
\end{itemize}
The even parameter $T$ and odd parameter $W$ do not obey any additional condition, and possess non-vanishing generators. The supertranslations are thus non-trivially realized. It follows that the asymptotic symmetry algebra at spatial infinity is given by the BMS algebra.  We refer to \cite{Henneaux:2018hdj,Henneaux:2019yax} for the details.

\section{Electric Carrollian limit of Einstein gravity with twisted parity conditions}
\label{sec:ElectricCarrolEinstein}

\subsection{Magnetic limit with diffeomorphism-twisted parity conditions}
The magnetic Carroll contraction of the boundary conditions given above takes exactly the same form, because the leading orders of the transformation of the fields under surface deformations coincide with the Einstein case.   Indeed, one now has
\begin{eqnarray}
\delta_{\xi,\xi^i} g_{ij}&=&\mathcal{L}_{\xi}g_{ij}\,, \label{eq:TransfCarrollMG}\\
\delta_{\xi,\xi^i} \pi^{ij}&=&-\frac{\xi}{\sqrt{g}}\left(R^{ij}-\frac{1}{2}g^{ij}R\right)
+\sqrt{g}\left(\nabla^i\nabla^j \xi-g^{ij}\nabla_m\nabla^m\xi\right)+\mathcal{L}_{\xi}\pi^{ij}\,,\label{eq:TransfCarrollMPi}
\end{eqnarray}
and one thus sees that for $\xi$ and $\xi^k$ of order one, the leading terms of (\ref{eq:TransfCarrollMG})-(\ref{eq:TransfCarrollMPi}) and (\ref{eq:TransfEinsteinG})-(\ref{eq:TransfEinsteinPi}) coincide.  The conditions on the leading terms of the metric and the conjugate momentum at spatial infinity are thus naturally taken to be exactly the same as in (\ref{eq:TwistedE1})-(\ref{eq:TwistedE2}).

Because the terms being dropped in the Hamiltonian constraints are algebraic, the analysis of the surface terms in the canonical generators proceeeds as in the Einstein theory. One can in particular easily verify that Carroll boosts are integrable when $\xbar \lambda_A = 0$ and that the asymptotic symmetry algebra is $\cC$-BMS, as found in \cite{Perez:2021abf} for the boundary conditions of \cite{Henneaux:2018cst}.

The twisted parity conditions produce therefore the expected result in the magnetic contraction. 
The electric Carrollian limit, on the other hand, opens new possibilities and we focus on it in the rest of this section.

\subsection{Electric limit: Action principle and transformation laws}\label{subsec:E-Carr-Action}
The Hamiltonian action principle of the electric Carrollian theory of gravity in four dimensions is given by \cite{Henneaux:1979vn}, \cite{Henneaux:2021yzg}
\be
S^E[g_{ij},\pi^{ij},N,N^i]=\int dt \left[\int d^{3}x \left(\pi^{ij}\dot{g}_{ij}-N \mathcal{H}^E- N^{i} \mathcal{H}^E_{i}\right)-B^E_{\infty}\right]\,.
\ee
Variation with respect to the lapse $N$ and shift $N^i$ functions enforces again the Hamiltonian and momentum constraints, which reads in the electric contraction, 
\begin{align}
\mathcal{H}^E & =\frac{1}{\sqrt{g}}\left(\pi^{ij}\pi_{ij}-\frac{\pi^{2}}{2}\right)\approx 0\,,\label{eq:HamE4D}\\
\mathcal{H}^E_{i} & =-2\nabla^{j}\pi_{ij}\approx 0 \,.\label{eq:MomE4D}
\end{align}
These constraints obey the ``zero-signature'' deformation algebra \cite{Teitelboim:1978wv}, \cite{Henneaux:1979vn}
\begin{eqnarray}
\{\cH ^E(x),\cH ^E(x')\}&=&0\,,\\
\{\cH ^E(x),\cH ^E_i(x')\}&=&\cH ^E(x)\partial_i\delta(x,x')\,,\\
\{\cH ^E_i(x),\cH ^E_j(x')\}&=&\cH ^E_i(x')\partial_j\delta(x,x')+\cH ^E_j(x)\partial_i\delta(x,x')\,.
\end{eqnarray}

Variation of the action with respect to the momentum $\pi^{ij}$ and the 3-dimensional metric $g_{ij}$ gives the following equations of motion
\begin{align}
 \dot{g}_{ij} & =\frac{2 N}{\sqrt{g}}\left(\pi_{ij}-\frac{1}{2}g_{ij}\pi\right)+N_{i|j}+ N_{j|i}\,,\label{eq:doth-elec4D}\\
\dot{\pi}^{ij} & =\frac{N}{2\sqrt{g}}g^{ij}\left(\pi^{mn}\pi_{mn}-\frac{\pi^{2}}{2}\right)
 -\frac{2N}{\sqrt{g}}\left(\pi^{im}\pi_{m}^{j}-\frac{1}{2}\pi^{ij}\pi\right)+\mathcal{L}_{N}\pi^{ij}\,.\label{eq:dotp-elec4D}
\end{align}

The infinitesimal transformation laws of the canonical pair $(g_{ij},\pi^{ij})$ under diffeomorphisms generated by $\left(\xi^{\perp}\equiv\xi,\xi^{i}\right)$ are now
\begin{align}
\delta_{\xi,\xi^i} g_{ij} & =\frac{2 \xi}{\sqrt{g}}\left(\pi_{ij}-\frac{1}{2}g_{ij}\pi\right)+\xi_{i|j}+ \xi_{j|i}\,,\label{eq:dh-elec4D}\\
\delta_{\xi,\xi^i} \pi^{ij} & =\frac{\xi}{2\sqrt{g}}g^{ij}\left(\pi^{mn}\pi_{mn}-\frac{\pi^{2}}{2}\right)
 -\frac{2\xi}{\sqrt{g}}\left(\pi^{im}\pi_{m}^{j}-\frac{1}{2}\pi^{ij}\pi\right)+\mathcal{L}_{\xi}\pi^{ij}\,.\label{eq:dp-elec4D}
\end{align}

One should note that in (\ref{eq:dotp-elec4D}) and (\ref{eq:dp-elec4D}), the terms involving $\xi$ decay at least as $r^{-3}$ at infinity, even for parameters $\xi$ that blow up as $r$,  so that the leading order $\xbar \pi^{ij}$ is invariant under normal hypersurface deformations. 

\subsection{Boundary conditions in the electric Carroll limit} \label{subsec:twistedCarroll}
In view of the remark just made, the Carroll contraction in the electric case of the twisted boundary conditions yield \begin{align}
	\xbar h_{rr}&= \text{even}\,,\\
	\xbar \lambda_A &= (\xbar \lambda_A)^{\text{odd}}+\xbar D_A \zeta_r-\xbar \zeta_A\,,\\ 
    \xbar h_{AB} &= (\xbar h_{AB})^{\text{even}}+\xbar D_A \xbar \zeta_B+ \xbar D_B \xbar \zeta_A+2\xbar g_{AB}\zeta_r\,,\\ 
    \xbar \pi^{rr}&= (\xbar \pi^{rr})^{\text{odd}}\,,\\
    \xbar \pi^{rA}&= (\xbar \pi^{rA})^{\text{even}}\,,\\
    \xbar \pi^{AB}&= (\xbar \pi^{AB})^{\text{odd}}\,,
\end{align}
where the leading order of the conjugate momentum is now strictly even or odd since the corresponding improper gauge terms are of lower order.

The contraction of the twisted boundary conditions also implies $\xbar \lambda_A =0$, but this condition turns out not to be necessary in the limit.  This condition can be consistently avoided, leading to a bigger symmetry group. 

\subsection{Asymptotic symmetries of the electric Carroll limit}
\label{subsec:Asymptotic conditions4D}
We now give the transformation laws of the leading orders of the fields under Carrollian diffeomorphisms with asymptotic form
\begin{eqnarray}\label{eq:diff}
\xi&=&br +T+\mathcal{O}\left(r^{-2}\right)\,,\\
\xi^r&=&W+\mathcal{O}\left(r^{-1}\right)\,,\\
\xi^A&=&Y^A+\frac{1}{r}I^A+\mathcal{O}\left(r^{-2}\right)\,,
\end{eqnarray}
where $b$ describes Carrollian boosts, $Y^A$ spatial rotations, and $T$ and $W$ stand for Carrollian supertranslations.  The parameter $I^A$ also describes spatial supertranslations and its precise role will depend on whether we impose $\xbar \lambda_A =0$ or not (see below). 


We find that the transformations of the leading orders under space-like diffeomorphisms read
\begin{align}
\delta_{\xi^i} \xbar h_{rr} & =Y^{A}\partial_{A}\xbar h_{rr} \,,\label{eq:dhrr-1-2}\\
\delta_{\xi^i}\overline{\lambda}_{A} & =\mathcal{L}_{Y}\overline{\lambda}_{A}+\xbar D_A W-I_A\,,\\
\delta_{\xi^i}\overline{h}_{AB} & =\mathcal{L}_{Y}\overline{h}_{AB}+2\left(\overline{D}_{(A}I_{B)}+\overline{g}_{AB}W\right)\,.
\end{align}
We also find that the leading orders of the momentum transform as
\begin{align}
\delta_{\xi^i}\xbar \pi^{rr}&=  \partial_{A}\left(Y^{A}\xbar \pi^{rr}\right)\,,\\
\delta_{\xi^i}\xbar \pi^{rA}&=  \partial_B\left(Y^B\xbar \pi^{rA}\right)-\partial_B Y^A \xbar \pi^{rB}\,,\\
\delta_{\xi^i}\xbar \pi^{AB}&=  \partial_C\left(Y^C \xbar \pi^{AB}\right)-\partial_CY^A\xbar \pi^{CB}-\partial_CY^A\xbar \pi^{CA}\,,
\end{align}
while the subleading order $\pi^{(2)rA}$, which appears in the expression of the charges as in the Einstein theory, transforms as
\begin{align}
 \delta_{\xi^i}\pi^{(2)rA}&= \partial_B\left(Y^B \pi^{(2)rA}\right)-\partial_B Y^A \pi^{(2)rB} \nonumber\\
&\quad +\partial_B\left(I^B\xbar \pi^{rA}\right)-\partial_B I^A \xbar \pi^{rB}+I^A \xbar \pi^{rr}-\partial_B W \xbar \pi^{AB}-W\xbar \pi^{rA}\,.
\end{align}

Similarly, we find that the transformation law under  time-like deformations are
\begin{align}
\delta_{\xi}\xbar h_{rr} & =\frac{b}{\sqrt{\xbar g}}\Big(\xbar \pi^{rr}-\xbar \pi^A_A\Big)\,,\label{eq:dhrr-1-1-1}\\
\delta_{\xi}\overline{\lambda}_{A} & =\frac{2b}{\sqrt{\xbar g}}\xbar \pi^{r}_A\,,\\
\delta_{\xi}\xbar h_{AB} & =\frac{2b}{\sqrt{\xbar g}}\Big[\xbar \pi_{AB}-\frac{1}{2}\xbar g_{AB}(\xbar \pi^{rr}+\xbar \pi^A_A)\Big]\,.
\end{align}
The leading orders of the momentum do not transform under time-like deformations, 
\be
\delta_{\xi}\xbar \pi^{rr} =
\delta_{\xi^i}\xbar \pi^{rA}=  \delta_{\xi^i}\xbar \pi^{AB} = 0\, , 
\ee
but one finds for $\pi^{(2)rA}$, \begin{equation}
\delta_{\xi}\pi^{(2)rA}=-\frac{2b}{3\sqrt{\xbar g}}\Big(2\xbar \pi^{rr}\xbar \pi^{rA}+3\xbar \pi^{r}_B \xbar \pi^{AB}-\xbar \pi^{rA}\xbar \pi^B_B \Big)\,.
\end{equation}

\subsubsection{Canonical generator}
Because the symplectic structure takes the standard form $\int d^3x \, d_V \pi^{ij} \wedge d_V g_{ij}$, the canonical generator of the asymptotic symmetries is simply obtained through the approach of \cite{Regge:1974zd}. The surface integral in this case reads
\begin{equation}\label{eq:dQE}
\delta Q^E_{\xi}[g_{ij},\pi^{ij}]=\oint d^2x\Big(2\xi^i\delta\pi^r_i-\xi^r\delta g_{jk}\pi^{jk} \Big)\,. 
\end{equation}
Replacing the asymptotic conditions in \eqref{eq:dQE}, we obtain
\begin{align}
\delta Q^E_{\xi}[g_{ij},\pi^{ij}]&=r \oint d^2x \Big(2Y^A\delta \xbar \pi^r_B\Big) \\
&\quad+ \oint d^2x\Big[2Y^A\delta \left(\pi^{(2)r}_A+\xbar h_{AB}\xbar \pi^{rB}+\xbar \lambda_A\xbar \pi^{rr}\right)+2I^A\delta \xbar \pi^r_A+2W\delta \xbar \pi^{rr}\Big] \,.
\end{align}
The variation of the canonical generator has a linear divergent term. However, this is equal to zero by virtue of the asymptotic constraint $\xbar D_A\xbar \pi^{AB}+\xbar \pi^{rA}=0$, and using the fact that $Y^A$ is the Killing vector of the 2-sphere at infinity. Then, the variation of the charge is given by
 \begin{align}
 \label{eq:dQEBis}
    \delta Q^E_{\xi}[g_{ij},\pi^{ij}]&  = \oint d^2x\Big[2Y^A \delta\left(\pi^{(2)r}_A+\xbar h_{AB}\xbar \pi^{rB}+\xbar \pi^{rr}\xbar \lambda_A\right)+2I^A \delta \xbar \pi^r_A+2W \delta \xbar \pi^{rr}\Big] \,.
\end{align}

To proceed further, we need to distinguish two cases, according to whether the additional  asymptotic condition $\xbar \lambda_A=0$ is imposed as in the Einstein theory, or is not imposed, as it turns out to be possible.

\subsubsection{Asymptotic symmetry algebra - Case with $\xbar \lambda_A=0$: $iso_3$-BMS algebra}

When $\xbar \lambda_A=0$, the parameters $I_A$ are not independent.  Indeed, preservation of the condition $\xbar \lambda_A=0$ requires one to add the following correcting gauge transformation when one performs a Carroll boost and a spatial supertranslation parametrized by $W$,
\begin{equation}
    \xi^r = \mathcal O\left(r^{-2}\right)\,, \qquad \xi^A = \frac 1 r (I^A_{(b)} + I^A_{(W)} )+ \mathcal O\left(r^{-2}\right)\,,
\end{equation}
with
\begin{equation}
    I^A _{(b)}= \frac {2b}{\sqrt {\xbar g}}\xbar \pi^{rA} \qquad I^A_{(W)} = \xbar D_A W\,,
\end{equation}
so that the $I_A$'s are entirely determined by $b$ and $W$. 

Substituting this expression for $I_A$ in $ \delta Q^E_{\xi}[g_{ij},\pi^{ij}]$, one finds after integration in phase space that the canonical generator is equal to
\begin{equation}
Q^E_{\xi}=\frac{1}{2}b_{ij}M^{ij}+Q_{W}\,,
\end{equation}
where
\begin{eqnarray}
M^{ij}&=&\oint d^2x\, 2x^{[i} e^{j]A} \left(\pi^{(2)r}_A+\xbar h_{AB}\xbar \pi^{rB}+\xbar \pi^{rr}\xbar \lambda_A\right)\,,\\
Q_W&=&\oint d^2 x\,W\mathcal{W}\,,
\end{eqnarray}
with $\mathcal{W}=2(\xbar \pi^{rr}-\xbar \pi^A_A)$. The surface term accompanying the boost is given by
\be \oint d^2x \,\frac{2n^i}{\sqrt{\xbar g}} \xbar \pi^{rA}\xbar \pi_{rA}\,,
\ee
but this integral is equal to zero due to the parity conditions.  It follows that the boosts are also proper gauge transformations for the diffeomorphism-twisted parity conditions, as in \cite{Perez:2021abf}.  Furthermore, it can be checked that  only the odd part of $W$ is an improper gauge transformation. 

The parameters obey the following transformation laws
\begin{eqnarray}
\hat{Y}^A&=&Y_1^{B}\partial_{B}Y_2^A-(1\leftrightarrow2)\,,\\
\hat W&=& Y_1^B\partial_B W_2-(1\leftrightarrow2)\,,
\end{eqnarray}
leading to the brackets
\begin{align}
\{M^{ij},M^{kl}\}&=\frac{1}{2}\Big(\delta^{jk}M^{il}-\delta^{ik}M^{jl}+\delta^{li}M^{jk}-\delta^{lj}M^{ik}\Big)\,, \\
\{M^{ij},\mathcal W(\theta,\phi)\}&=-\xbar D_B(x^{[i}e^{j]B} \mathcal W(\theta,\phi))\,.
\end{align}
This is the $iso(3)$-BMS algebra.

\subsubsection{Asymptotic symmetry algebra - Case with $\xbar \lambda_A \neq 0$: Extended $iso(3)$-BMS algebra}

Unlike what happens in the magnetic Carrollian theory, the expression (\ref{eq:dQEBis}) is integrable for arbitrary $Y^A$'s, $I^A$'s and $W$'s.  Thus, there is no need to impose the restriction $\xbar \lambda_A=0$ which was necessary in that case  (and also in Einstein gravity) to ensure integrability.  We thus lift this restriction and allow configurations with  $\xbar \lambda_A \neq 0$.

The electric theory is then invariant under the semi-direct product of spatial rotations and the Abelian set of ``generalized'' spatial Carrollian supertranslations parametrized by $W$ and $I^A$, with canonical generator that can be written as
\begin{equation}
Q^E_{\xi}=\frac{1}{2} b_{ij}M^{ij}+Q_W+Q_I\,,
\end{equation}
where
\begin{equation}
Q_W=\oint d^2x\, W \cW \,,\qquad  Q_I=\oint d^2x\,  I^A \cW _A\,. 
\end{equation}
Here, $\mathcal W=2\xbar \pi^{rr}$ and $\mathcal W_A=2 \xbar \pi^r_A$.  Because $\xbar \pi^{rr}$ is odd and $\xbar \pi^r_A$ even, only odd parameters $W$ and even parameters $I^A$, which are otherwise  arbitrary functions and vectors on the 2-sphere, define improper gauge transformations.

The transformation laws of the parameters are given by
\begin{eqnarray}
\hat{Y}^A&=&Y_1^{B}\partial_{B}Y_2^A-(1\leftrightarrow2)\,,\\
\hat W&=& Y_1^B\partial_B W_2-(1\leftrightarrow2)\,,\\
\hat I^{A}&=&Y_1^B\partial_B I^{A}_2-I^{B}_2\partial_BY_1^A -(1\leftrightarrow2) \,,
\end{eqnarray}
and hence, the nonvanishing brackets of the generators read
\begin{align}
\{M^{ij},M^{kl}\}&=\frac{1}{2}\Big(\delta^{jk}M^{il}-\delta^{ik}M^{jl}+\delta^{li}M^{jk}-\delta^{lj}M^{ik}\Big)\,,\\
\{M^{ij},\mathcal W (\theta,\phi)\}&=-\xbar D_B(x^{[i}e^{j]B} \mathcal W(\theta,\phi))\,,\\
\{M^{ij},\cW_A (\theta,\phi)\}&=\xbar D_B(x^{[i}e^{j]B} \cW_A(\theta,\phi))-x^{[i}\xbar D_Ae^{j]B}\cW_B (\theta,\phi)\,.
\end{align}
This is the extended $iso(3)$-BMS algebra.

This enlargment of the $iso(3)$-BMS algebra cannot be obtained as a contraction of the BMS$_4$ algebra since it contains more spatial supertranslations.  It is somewhat reminiscent of the Spi group of \cite{Ashtekar:1978zz}, but the parity of the supertranslations are restricted here.

\section{Asymptotic symmetries of the Yang-Mills field in flat Carroll spacetime}
\label{sec:ElectricCarrollYM}
Before turning to the coupled Einstein-Yang-Mills system in the Carroll limit, we first consider the asymptotic symmetries of the  Yang-Mills field on a flat Carroll background. We start with a brief overview of the asymptotic structure of the Yang-Mills theory in Minkowski space. We then proceed to the analysis of the electric and magnetic Carrollian limits.

\subsection{Brief overview of boundary conditions in Yang-Mills theory}
The asymptotic structure of Yang-Mills theory has  been rigorously studied in \cite{Tanzi:2020fmt}. Here, we give a brief summary of their results. 

The Hamiltonian action for the Yang-Mills theory on a Minkowski background in four spacetime dimensions reads
\begin{equation}
S_{\text{YM}}[A_i,\pi^{i},A_0]=\int dt d^3 x \left(\pi^i_a \dot A^a_i-\mathcal H +A^a_0 D _i \pi^i_a \right)\,,
\end{equation}
where $\pi^i_a$  is the conjugate momentum to the non-Abelian gauge field $A_i=A^a_i T_a$. This canonical pair takes values on some compact semi-simple Lie algebra $[T_a, T_b]= f^c_{\;\;ab} T_c$. By using the inverse of the invariant metric, one can raise the internal index $a$ of $\pi^i_a$ to form an object $\pi^i = \pi^{ia} T_a$ that transforms as the vector potential in the adjoint representation.\footnote{Here, we follow the conventions of \cite{Tanzi:2020fmt}. The inner product is given by $S_{ab}=-\text{tr}(T_a T_b)$, where $S^{ab}S_{bc}=\delta^a_c$. Latin indices $a,b,c,\dots$ refer to the internal Lie algebra and  are lowered and raised with $S_{ab}$ and its inverse $S^{bc}$, respectively.} The variation of the action with respect to the Lagrange multiplier $A^a_0$ enforces the Gauss constraint
\begin{equation}
\mathcal G_a=D _i \pi^i_a\approx 0\,,
\end{equation}
where the covariant derivative is defined as $D_i X^a = \partial_i X^a+\alpha\,f^a_{\;\;bc} A^b_i X^c$, with $\alpha$ the Yang-Mills coupling constant. The Hamiltonian density is given by
\begin{equation}
\mathcal H =\frac{1}{2}\pi^a_i \pi_a^i+ \frac{1}{4}F^a_{ij}F_a^{ij} \,,
\end{equation}
while the momentum density reads
\be
\mathcal H_i =\pi^j_a \partial_i A^a_j-\partial_j(\pi^j_a A^a_i) \, .
\ee

Infinitesimal transformation laws of the fields under Poincar\'e and non-Abelian gauge symmetries, generated by $\xi^\mu=(\xi,\xi^i)$ and $\varepsilon=\varepsilon^a T_a$, read
\begin{align}
\delta_{\xi,\xi^i,\varepsilon} A_i^a&= \xi \pi_i^a+ \xi^j \partial_i A_j^a + \partial_i \xi ^j A_j^a - D_i\varepsilon^a\,,
\\
\delta_{\xi,\xi^i,\varepsilon} \pi^i_a&= -D_j(\xi F^{ij}_a) +\partial_j \left( \xi^j \pi ^i_a\right) - \partial_j \xi^i \pi^j_a + \alpha f^{c}_{\;\;ab}\pi^i_c\varepsilon^{b}\,.
\end{align}
In spherical coordinates, the vector fields defining the infinitesimal Poincar\'e transformations take the form
\begin{equation}
 \xi= br+T_0\,, \quad \xi^r=W_0\,, \quad \xi^A=Y^A+\frac{1}{r}\xbar D^A W_0\,,
\end{equation}
where $T_0$ stand for time translations $\partial_A T_0=0$ and $W_0$, which corresponds to the spatial translations, is a combination of the $l=1$ spherical harmonics and satisfies therefore the conditions $\xbar D_A \xbar D_B W_0+\xbar g_{AB}W_0=0$. The decay of the gauge parameter reads
\begin{equation}\label{eq:falloff-gauge}
\varepsilon=\xbar\varepsilon+\mathcal O(r^{-1})\,.
\end{equation}
The fall-off of the gauge field and its conjugate momentum is given by
\begin{align}
A_r &= \frac{\xbar A_r}{r} + \cO\left(r^{-2}\right)\,,\quad A_{A} = \xbar A_{A} + \cO \left(r^{-1}\right)\,,  \label{eq:falloff-A's}\\
\pi^r &= \xbar \pi^r  + \cO\left(r^{-1}\right)\,\quad \pi^{A} =\frac{\xbar \pi^{A}}{r} + \cO\left(r^{-2}\right)\,. \label{eq:falloff-pi's}
\end{align}

As in the case of electromagnetism \cite{Henneaux:2018gfi} and Einstein gravity \cite{Henneaux:2018hdj,Henneaux:2019yax}, the symplectic structure is logarithmically divergent.  This can be solved by imposing  parity conditions.  The authors of \cite{Tanzi:2020fmt} first propose the following set of strict parity conditions
\begin{equation}
   \xbar A_{r} \sim \xbar \pi^{A}=\text{even}\,, \quad \xbar A_A \sim \xbar \pi^{r}=\text{odd}\,,
\end{equation}
which ensures the finiteness of the symplectic structure. These parity conditions differ from those usually imposed in electromagnetism where $\xbar A_i$ is odd instead of being even, but possess two good features: (i) to leading order, the vector potential $A_i$ and the derivative operator $\partial_i$ possess same odd parity, so that $D_i = \partial_i - A_i$ has a definite parity (namely, also odd parity); and (ii) the $SU(2)$ Wu-Yang monopole solution of the pure Yang-Mills theory \cite{Wu:1967vp},
\be
A_i^a = \delta^{ab}\epsilon_{ibc} \frac{x^c}{r^2}
\ee
fulfills these parity conditions. 

Furthermore, these boundary conditions are Poincar\'e invariant and make the Lorentz boosts canonical.  They are also preserved by gauge transformations with gauge parameter that asymptotically tends to an even function on the 2-sphere, i.e., $\xbar \varepsilon=\text{even}$. However, the boundary term of the canonical generator of gauge transformations
\begin{equation}\label{eq:Gauge-generator}
G[\varepsilon] = \int d^3 x\, \varepsilon^a \, \mathcal G_a  -\oint d^2S_i \varepsilon^a \pi^i_a\,,
\end{equation}
is then equal to zero since $\varepsilon^a$ and  $\pi^i_a$ have opposite parity.  There is no improper gauge symmetry at all, contrary to what happens for electromagnetism.  The only symmetries are the Poincar\'e transformations, and these are rigid symmetries with a non-vanishing bulk integral.

The authors of \cite{Tanzi:2020fmt} relaxed then the boundary conditions by twisting the parity conditions 
as in \cite{Henneaux:2018gfi,Henneaux:2018hdj,Henneaux:2019yax}.
They imposed accordingly strict parity conditions up to an improper gauge transformation. Specifically, they proposed that the Yang-Mills field should behave asymptotically as
\begin{align}
\xbar A_r &=\xbar{U}^{-1}\xbar A_r^{\text{even}}\xbar{U}\,,\quad
\xbar A_{A} = \xbar{U}^{-1}\xbar A_{A}^{\text{odd}}\xbar{U}+\xbar{U}^{-1}\partial_{A}\xbar{U}\,,\label{eq:bc-A's}\\
\xbar \pi^r &=\xbar{U}^{-1}\xbar \pi^r_{\text{odd}}\xbar{U}\,,\qquad
\xbar \pi^{A}  = \xbar{U}^{-1}\xbar \pi^{A}_{\text{even}}\xbar{U}\,,\label{eq:bc-pi's}
\end{align}
where $\xbar{U}=\exp(-\xbar \phi^a \, T_a)$, where $\xbar \phi^a$ depends on the angles. They verified that these boundary conditions lead to a finite symplectic structure (by imposing a faster fall-off of the Gauss constraint $\mathcal G_a\sim r^{-4}$).  By construction,  these conditions are preserved by angle-dependent gauge symmetries with no definite parity and hence non-zero charges. However, the authors of \cite{Tanzi:2020fmt} also showed that there was a clash with Poincar\'e invariance, in the sense  that the symplectic structure is  not invariant under Lorentz boosts, which transforms into a non-trivial surface term.  While this problem can be cured in electromagnetism by adding boundary terms in the symplectic form \cite{Henneaux:2018gfi}, the same method does not work in the non-abelian case \cite{Tanzi:2020fmt}.  The difficulty is that the interaction term $A_i A_j$ in the curvature $F_{ij}$ is of the same order as the free term $\partial _i A_j$ and hence, the interactions cannot be neglected asymptotically.

The conclusion is therefore that there is no known set of asymptotic conditions at spatial infinity simultaneously consistent with Lorentz invariance and accommodating improper color gauge symmetries in the non-Abelian case.

\subsection{Electric Carrollian limit of Yang-Mills}

\subsubsection{Boundary conditions and asymptotic transformations} 
We now show  that the aforementioned difficulties do not appear in the electric Carrollian limit of Yang-Mills theory, which possesses an infinite-dimensional color symmetry group consistent with Carroll invariance. The situation in the magnetic Carrollian limit, on the other hand, is similar to the Lorentzian case, i.e., one easily verifies that it is not possible to have improper gauge transformations consistent with Carroll invariance. Thus, the asymptotic symmetry algebra is given in that case by the Carroll algebra only.  For that reason, from now on we focus on the electric limit, which has a richer structure.

The Hamiltonian action for the electric Carrollian limit of Yang-Mills theory reads \cite{Henneaux:2021yzg}
\begin{equation}
S^{E}_{\text{YM}}[A_i,\pi^{i},A_0]=\int dt d^3 x \left(\pi^i_a \dot A^a_i-\mathcal H^{E} +A^a_0 D _i \pi^i_a \right)\,,
\end{equation}
The Hamiltonian density is given  by
\begin{equation}\label{eq:Elec-densities}
\mathcal H^E =\frac{1}{2}\pi^a_i \pi_a^i \, ,
\end{equation}
and satisfies the relation $\{\mathcal H^E (x), \mathcal H^E(x')\} = 0$ characteristic of Carroll-invariant theories.  The momentum density and the Gauss constraint are unchanged.
The variation of the action with respect to the fields yields the following equations of motion
\begin{align}
 \dot{A}_i^a&=  \pi_i^a- D_iA ^a_0\,,\\
\dot{\pi}^i_a&= 0\,.
\end{align}

Carroll transformations are generated by the same surface-deformation vector fields as in the Lorentzian case, i.e., 
\begin{equation}
 \xi= br+T_0\,, \quad \xi^r=W_0\,, \quad \xi^A=Y^A+\frac{1}{r}\xbar D^A W_0\,,
\end{equation}
where $b$ is the Carrollian boost parameter, $Y^A$ is the Killing vector of the 2-sphere at infinity, the constant $T_0$ stands for time translations and $W_0$ corresponds to the spatial translations (satisfying the property $\xbar D_A \xbar D_B W_0+\xbar g_{AB}W_0=0$).   The action on the fields is however different because the Hamiltonian density is different, leading to variations of the fields under Carroll and gauge transformations that read
\begin{align}
\delta_{\xi,\xi^i,\varepsilon} A_i^a&= \xi \pi_i^a+ \xi^j \partial_i A_j^a + \partial_i \xi ^j A_j^a - D_i\varepsilon^a\,,
\\
\delta_{\xi,\xi^i,\varepsilon} \pi^i_a&= \partial_j \left( \xi^j \pi ^i_a\right) - \partial_j \xi^i \pi^j_a + \alpha f^{c}_{\;\;ab}\pi^i_c\varepsilon^{b}\,.
\end{align}
The notable difference lies in the transformation rule of the momenta $\pi^i_a$ under timelike deformations (boosts and time translations), which is now $\delta_{\xi} \pi^i_a = 0$. 

We adopt the same fall-off of the fields as in the previous section, i.e., \eqref{eq:falloff-A's}, \eqref{eq:falloff-pi's} and the set of twisted parity conditions \eqref{eq:bc-A's}, \eqref{eq:bc-pi's}. We also take the same asymptotic behaviour \eqref{eq:falloff-gauge} of the gauge parameter $\varepsilon$.  As we recalled, these boundary conditions make the symplectic form finite under the additional requirement that the Gauss constraint should decay as $D _i \pi^i_a\sim r^{-4}$.  It is direct to show that this  set of boundary conditions is preserved by both Carroll and  non-Abelian gauge symmetries behaving as in \eqref{eq:falloff-gauge}.

The transformation laws of the leading orders of the fields in the asymptotic expansion ($r \rightarrow \infty$) are given in spherical coordinates  by
\begin{eqnarray}
\delta_{\xi,\varepsilon} \xbar A^a_r&=& \frac{b\,\xbar \pi^{ar}}{\sqrt{\xbar g}}+Y^A \partial_A \xbar A^a_r+\alpha f^{a}_{\,\,\,bc}\xbar \varepsilon^b \xbar A^c_r\,,\\
\delta_{\xi,\varepsilon} \xbar A^a_A&=& \frac{b \,\xbar \pi^a_A}{\sqrt{\xbar g}}+Y^B \partial_B \xbar A^a_A + \partial_B Y^B\xbar A^a_A-D_A\bar\varepsilon^a\,,
\\
\delta_{\xi,\varepsilon} \xbar \pi^r_a &=& \partial_A \left(Y^A\xbar \pi^r_a\right)+\alpha f_{abc}\xbar \varepsilon^b \xbar \pi^{cr}\,,\\
\delta_{\xi,\varepsilon} \xbar \pi^A_a &=& \partial_B Y^B \xbar\pi_a^A - \partial_BY^A \xbar \pi_a^B +\alpha f_{abc}\xbar \varepsilon^b \xbar \pi^{c\,A}\,.
\end{eqnarray}

\subsubsection{Boosts are canonical transformations - Improper gauge transformations}
The great simplification that occurs in the electric Carroll contraction of Yang-Mills theory is that the boosts are now canonical transformations, even with the parity conditions twisted by a gauge transformation.  This is because there are no spatial derivatives in the energy density \eqref{eq:Elec-densities}. The twisted parity conditions are therefore compatible with Carroll invariance, while they were not (and could not be improved to become so) in the Lorentzian case.

As we have also explained above, the twisted parity conditions  \eqref{eq:bc-A's}, \eqref{eq:bc-pi's} are invariant under angle-dependent $\mathcal O(1)$ color gauge transformations, which are improper when they are odd (in a sense to make precise in the subsection below).  

The electric Carroll contraction of Yang-Mills theory accommodates consequently an infinite-dimensional angle-dependent color group without conflict with Carroll covariance.

\subsubsection{Asymptotic symmetry algebra: infinite-dimensional color group}
By applying the standard canonical methods, one then finds that the canonical generator of the asymptotic symmetries is given by
\begin{equation}
C^{E}_{\xi,\varepsilon}[A_{i},\pi^{i}]=\int d^3x \left(\xi\mathcal{H}^{E}+\xi^i \mathcal{H}_i+\varepsilon^a \mathcal{G}_a\right)+Q^{E}_{\xi,\epsilon}[A_{i},\pi^{i}]\,,
\end{equation}
where the surface term reads
\begin{equation}
Q^{E}_{\xi,\varepsilon}[A_{i},\pi^{i}]=\oint d^2x \left( -\xbar \varepsilon^a \xbar \pi^r_a +Y^A \xbar A^a_A \xbar \pi^r_a \right)\,.
\end{equation}
The generator of spatial rotations is the sum of both a non-vanishing bulk term $\int d^4x \xi^i \mathcal{H}_i$ (recall that $\mathcal{H}^{E}$ and $\ \mathcal{H}_i$ are not constrained to vanish when Carroll gravity is not included) and a non-vanishing surface integral $\oint d^2 xY^A \xbar A^a_A \xbar \pi^r_a$

A direct computation shows that the Poisson brackets of the canonical generators are given by
\begin{equation}
\Big\{ C^E_{\xi_1,\varepsilon_1}[A_{i},\pi^{i}],C^E_{\xi_2, ,\varepsilon_2}[A_{i},\pi^{i}]\Big\} =C^E_{\hat \xi, \hat \varepsilon}[A_{i},\pi^{i}]\,,
\end{equation}
where
\begin{eqnarray}
\hat \xi&=& \xi^i_1 \partial_i \xi_2-(1\leftrightarrow2)\,,\\
\hat \xi^i&=&\xi^j_1\partial_j \xi^i_2-(1\leftrightarrow2)\,,\\
\hat \varepsilon^a &=& \xi^i_1\partial_i \varepsilon_2^a+\frac{1}{2}f^a_{\,\,\,bc} \varepsilon^b_1 \varepsilon^c_2 -(1\leftrightarrow2)\,.
\end{eqnarray}
The symmetry algebra is the semi-direct sum of the Carroll algebra and an infinite-dimensional set of non-Abelian angle-dependent color charges. In order to write the algebra in a more explicit way, we can recast the canonical generator $C^{E}_{\xi,\varepsilon}[A_{i},\pi^{i}]$ as
\begin{equation}
C^{E}_{\xi,\varepsilon}=b_iB^i+a_0 E+\frac{1}{2}b_{ij}M^{ij}+a_i P^i+Q^{\text{YM}}_{\varepsilon}\,,
\end{equation}
where
\begin{eqnarray} 
B^i&=&\int d^3x x^i \cH^E\,,\\
E&=&\int d^3x \, \cH^E\,,\\
M^{ij}&=& \int d^3x\,2x^{[i}\cH^{j]}+\oint d^2 x \, x^{[i}e^{j]A}\xbar A^a_A\xbar \pi^r_a\,,\\ 
P^i&=& \int d^3x\,\cH^{i}\,,\\
Q^{\text{YM}}_{\epsilon}&=&\oint d^2x \xbar \epsilon^a \mathcal{T}_a\,, 
\end{eqnarray}
with $\mathcal{T}_a=-\xbar \pi^r_a$. Then, the non-vanishing brackets of the  asymptotic symmetry algebra are explicitly given by
\begin{align}
\{P^i,B^j\}&=\delta^{ij}E\,,\\
\{B^i,M^{jk}\}&= \frac{1}{2}\Big(\delta^{ik}B^j-\delta^{ij}B^k\Big)\,,\\
\{P^i,M^{jk}\}&= \frac{1}{2}\Big(\delta^{ik}P^j-\delta^{ij}P^k\Big)\,,\\
\{M^{ij},M^{kl}\}&=\frac{1}{2}\Big(\delta^{jk}M^{il}-\delta^{ik}M^{jl}+\delta^{li}M^{jk}-\delta^{lj}M^{ik}\Big)\,,\\
\{M^{ij},\mathcal T_a(x^A)\}&=-\xbar D_B(x^{[i}e^{j]B} \mathcal T_a(x^A))\,,\\
\{\mathcal T_a(x^A_1),\mathcal T_b(x^A_2)\}&=f^c_{\,\,\,ab}\mathcal T_c(x^A_1)\delta^{(2)}(x^A_1-x^A_2)\,.
\end{align}

Note that the generators $\mathcal{T}_a=-\xbar \pi^r_a$ of the improper color gauge transformations, which form  a centerless Kac-Moody algebra, obey a twisted parity condition and thus they are not independent functions on the sphere.

\section{Carrollian limits of Einstein-Yang-Mills system}
\label{sec:ElectricCarrollEinsteinYM}

In this section, we proceed to combine our previous results, which can be  done easily. One can take different limits in the Einstein and Yang-Mills sectors of the coupled theory, which would lead to different consistent Carrollian limits of the Einstein-Yang-Mills system. We will consider first, and somewhat in detail, the theory that results from taking the magnetic Carrollian limit for gravity and the electric Carrollian limit for Yang-Mills, since it is the one that admits the most interesting asymptotic structure.  We will then make a summary of the asymptotic symmetry algebras associated to the remaining Carrollian limits.

\subsection{Magnetic Carrollian gravity coupled to the electric limit of the Yang-Mills field}

The Hamiltonian action for this Carrollian limit of the Einstein-Yang-Mills system reads
\begin{equation}
S=\int dt \left[\int d^{3}x \left(\pi^{ij}\dot{g}_{ij}+\pi^i_a\dot{A}^a_i-N \mathcal{H}- N^{i} \mathcal{H}_{i}+A^a_0\mathcal{G}_a\right)-B_{\infty}\right]\,.
\end{equation}
The variation of the action with respect to the Lagrange multipliers $N$, $N^i$ and $A^a_0$ yields the following constraints
\begin{eqnarray}
\cH &=& -\sqrt{g}R+\frac{1}{2\sqrt{g}}\pi^a_i \pi_a^i\approx 0\,,\\
\cH_i &=&-2\nabla^{j}\pi_{ij}+\pi^j_a \partial_i A^a_j-\partial_j(\pi^j_a A^a_i)\approx 0\,,\\
\cG_a &=& D_i\pi^i_a\approx 0\,,
\end{eqnarray}
where the covariant derivative $D_i$ is the complete spacetime ($\nabla_i$) and gauge covariant derivative,   $D_i X^a = \nabla_i X^a+\alpha\,f^a_{\;\;bc} A^b_i X^c$. 

The constraints are all first-class and satisfy the following algebra
\begin{eqnarray}
\{\cH (x),\cH (x')\}&=&0\,,\\
\{\cH (x),\cH _i(x')\}&=&\cH (x)\partial_i\delta(x,x')\,,\\
\{\cH _i(x),\cH _j(x')\}&=&\cH _i(x')\partial_j\delta(x,x')+\cH _j(x)\partial_i\delta(x,x')\,,\\
\{\cG_a (x),\cH (x')\}&=&0\,,\\
\{\cG_a (x),\cH_i (x')\}&=&\cG_a (x)\partial_i\delta(x,x')\,,\\
\{\cG_a (x),\cG_b (x')\}&=&f^c_{\,\,\,ab}\cG_c(x)\delta(x,x')\,.
\end{eqnarray}
The infinitesimal transformation laws of the fields generated by Carrollian diffeomorphisms $\left(\xi^{\perp}\equiv\xi,\xi^{i}\right)$ and non-Abelian gauge transformation with parameter $\varepsilon^a$ read
\begin{align}
\delta_{\xi,\xi^i} g_{ij} & =\mathcal{L}_{\xi}g_{ij}\,,\label{eq:dh-mag}\\
\delta_{\xi,\xi^i}\pi^{ij} & =-\xi\sqrt{g}\left(R^{ij}-\frac{1}{2}g^{ij}R\right)+\frac{\xi}{2\sqrt{g}}\left(\pi^{ai}\pi^j_a-\frac{1}{2}g^{ij}\pi^a_k \pi_a^k\right)\\
&\quad+\sqrt{g}\left(\xi^{|ij}-g^{ij}\xi_{\quad|m}^{|m}\right)+\mathcal{L}_{\xi}\pi^{ij}\,,\label{eq:dp-mag}\\
\delta_{\xi,\xi^i,\varepsilon} A_i^a&= \frac{\xi}{\sqrt{g}} \pi_i^a+ \xi^j \partial_i A_j^a + \partial_i \xi ^j A_j^a - D_i\varepsilon^a\,,
\\
\delta_{\xi,\xi^i,\varepsilon} \pi^i_a&= \partial_j \left( \xi^j \pi ^i_a\right) - \partial_j \xi^i \pi^j_a + \alpha f^{c}_{\;\;ab}\pi^i_c\varepsilon^{b}\,.
\end{align}

We take as asymptotic conditions:
\begin{itemize}
\item for gravity, the twisted parity conditions \eqref{eq:TwistedE1}-\eqref{eq:TwistedE2}, with the additional condition $\xbar \lambda_A = 0$  (which implies  $\xbar \zeta_A=\xbar D_A\zeta_r$);
\item for the Yang-Mills field, the twisted conditions \eqref{eq:falloff-A's},  \eqref{eq:falloff-pi's}, \eqref{eq:bc-A's} and \eqref{eq:bc-pi's}.
\end{itemize}
Additionally, we  also assume the faster fall-off for the constraints
\begin{equation}
\mathcal{H}= \mathcal{O}(r^{-2})\,,\quad \mathcal{H}_r=\mathcal{O}(r^{-2})\,, \quad \mathcal{H}_A=\mathcal{O}(r^{-1})\,,\quad \mathcal{G}_a=\mathcal{O}(r^{-4})\,.
\end{equation}

The asymptotic Killing vectors and the non-Abelian gauge parameters that preserve the fall-off of the fields read
\begin{eqnarray}\label{eq:diff-EYM1}
\xi &=&br +T-\frac{1}{2}b\xbar h+\mathcal{O}\left(r^{-2}\right)\,,\\
\xi^r &=&W+\mathcal{O}\left(r^{-1}\right)\,,\\
\xi^A &=&Y^A+\frac{1}{r}\xbar D^A W+\mathcal{O}\left(r^{-2}\right)\,,\\
\varepsilon &=&\xbar \varepsilon+\mathcal O(r^{-1})\,,
\end{eqnarray}
where $\xbar D_A \xbar D_B b+\xbar g_{AB}b=0$, $\xbar D_A Y_B+\xbar D_B Y_A=0$ and where the boost-dependent $\mathcal O(1)$ compensating term $-\frac{1}{2}b\xbar h$ is included  in order to make the boost charges integrable, as in Einstein gravity (the details of this derivation can be found in \cite{Henneaux:2018cst,Henneaux:2019yax,Perez:2021abf}).

These transformations are canonical and hence define asymptotic symmetries. Their canonical generators are again obtained through the standard canonical procedure, and explicitly given by
\begin{equation}
G_{\xi,\epsilon}[g_{ij},\pi^{ij};A_{i},\pi^{i}]=\int d^3x \left(\xi\mathcal{H}+\xi^i \mathcal{H}_i+\epsilon^a \mathcal{G}_a\right)+Q_{\xi,\epsilon}[g_{ij},\pi^{ij};A_{i},\pi^{i}]\,
\end{equation}
where the boundary terms read
\begin{align}
Q_{\xi,\epsilon}&=\oint d^2x \Big[ Y^A(2\xbar \pi^{(2)r}_A+\xbar A^a_A \xbar \pi^r_a)+2\sqrt{\xbar g}\,T\,\xbar h_{rr}+2W(\xbar \pi^{rr}-\xbar \pi^A_A)\nonumber\\
&\quad \qquad \qquad+\sqrt{\xbar g}\,b\Big(2k^{(2)}+\frac{1}{2}\xbar h_{rr} \xbar h+\frac{1}{4}\xbar h^2 +\frac{1}{4}\xbar h^A_B \xbar h^B_A\Big)-\xbar \varepsilon^a \xbar \pi^r_a\Big]\,. 
\end{align}
with
\begin{align}
\xbar k & =\frac{1}{2}\xbar h+\xbar h_{rr}\,,\\
\xbar k^{(2)} & =h_{rr}^{(2)}+h^{(2)}-\frac{1}{4}\xbar h_{rr}\xbar h-\frac{1}{2}\xbar h_{B}^{A}\xbar h_{A}^{B}-\frac{3}{4}\xbar h_{rr}^{2}+\xbar D_{A}h_{r}^{(2)A}\,,\nonumber \\
\xbar\pi_{A}^{(2)r} & =\pi_{A}^{(2)r}+\xbar h_{AB}\xbar\pi^{rB}\,.
\end{align}

The brackets of the canonical generators are given by
\begin{equation}
\Big\{ G_{\xi_1,\varepsilon_1}[g_{ij},\pi^{ij};A_{i},\pi^{i}],G_{\xi_2,\varepsilon_2}[g_{ij},\pi^{ij};A_{i},\pi^{i}]\Big\} =G_{\hat \xi, \hat \varepsilon}[g_{ij},\pi^{ij};A_{i},\pi^{i}]\,,
\end{equation}
where the hatted parameters of the commutator transformation $(\hat \xi, \hat \varepsilon)$ are
\begin{eqnarray}
\hat{Y}^A&=&Y_1^{B}\partial_{B}Y_2^A-(1\leftrightarrow2)\,,\\
\hat b&= &Y_1^B\partial_B b_2-(1\leftrightarrow2)\,,\\
\hat T&=& Y_1^B\partial_B T_2-3b_1 W_2-\partial_A b_1 \xbar D^AW_2-b_1\xbar D_A\xbar D^AW_2-(1\leftrightarrow2)\,,\\
\hat W&=& Y_1^B\partial_B W_2-(1\leftrightarrow2)\,,\\
\hat \varepsilon^a &=& \xi^i_1\partial_i \varepsilon_2^a+\frac{1}{2}f^a_{\,\,\,bc} \varepsilon^b_1 \varepsilon^c_2 -(1\leftrightarrow2)\,.
\end{eqnarray}
These transformations define the Carroll-BMS algebra $\mathcal C$-BMS endowed with an infinite set of color charges.

Writing the canonical charges as
\begin{equation}
Q_{\xi,\epsilon}=b_iB^i+\frac{1}{2}b_{ij}M^{ij}+Q_{W}+Q_{T}+Q^{\text{YM}}_{\epsilon}\,,
\end{equation}
where
\begin{align}
&B^i=\oint d^2x \sqrt{\xbar g}\,n^i\Big(2k^{(2)}+\frac{1}{4}\xbar h^2 +\frac{1}{4}\xbar h^A_B \xbar h^B_A\Big) \,,\\
& M^{ij}=\oint d^2x x^{[i}e^{j]A}\Big(\xbar \pi^{(2)r}_A+\frac{1}{2}\xbar A^a_A \xbar \pi^r_a\Big)\,,\\
&Q_{T}=\oint d^2x \sqrt{\xbar g }\,T\, \mathcal T \,,\quad Q_{W}=\oint d^2x \,W\, \mathcal W\,,\\
&Q^{\text{YM}}_{\epsilon}=\oint d^2x\, \xbar \epsilon^a \mathcal{T}_a\,, 
\end{align}
with
\begin{equation}
\mathcal{T}=2\xbar h_{rr}\,,\qquad \mathcal{W}=2(\xbar \pi^{rr}-\xbar \pi^A_A)\,,\qquad\mathcal{T}_a=-\xbar \pi^r_a\,,
\end{equation}
we then find that the brackets of the Carroll boosts and spatial rotations are given by
\begin{align}
\{B^i,B^j\}&=0\,,\\
\{B^i,M^{jk}\}&= \frac{1}{2}\Big(\delta^{ik}B^j-\delta^{ij}B^k\Big)\,,\\
\{M^{ij},M^{kl}\}&=\frac{1}{2}\Big(\delta^{ik}M^{jl}-\delta^{il}M^{jk}-\delta^{jk}M^{il}+\delta^{jl}M^{ik}\Big)\,.
\end{align}
The brackets of the time and spatial Carrollian supertranslations and the non-Abelian charges with Carroll boosts and spatial rotations read
\begin{align}
\{B^i,\mathcal W(x^A)\}&=-3n^i \mathcal T-\partial_A n^i \xbar D^A \mathcal T-n^i \xbar \triangle \mathcal T\,,\\
\{B^i,\mathcal T(x^A)\}&=0\,,\\
\{B^i,\mathcal T_a(x^A)\}&=0\,,\label{eq:boost-gauge}\\
\{M^{ij},\mathcal W(x^A)\}&=-\xbar D_B(x^{[i}e^{j]B} \mathcal W(x^A))\,,\\
\{M^{ij},\mathcal T(x^A)\}&=-\xbar D_B(x^{[i}e^{j]B} \mathcal T(x^A))\,,\\
\{M^{ij},\mathcal T_a(x^A)\}&=-\xbar D_B(x^{[i}e^{j]B} \mathcal T_a(x^A))\,.
\end{align}
Finally, the bracket between the non-Abelian charges is given by
\begin{align}
\{\mathcal T_a(x^A_1),\mathcal T_b(x^A_2)\}&=f^c_{\,\,\,ab}\mathcal T_c(x^A_1)\delta^{(2)}(x^A_1-x^A_2)\,.
\end{align}

\subsection{Magnetic Carrollian gravity coupled to the magnetic limit of the Yang-Mills field}
In this case, the color charges vanish because of the strict set of parity conditions (of \cite{Tanzi:2020fmt}),
\begin{equation}
   \xbar A_{r} \sim \xbar \pi^{A}=\text{even}\,, \quad \xbar A_A \sim \xbar \pi^{r}=\text{odd}\,,
\end{equation} 
that one must impose on the Yang-Mills field in order to render Carroll boosts canonical transformations. Thus, the asymptotic symmetry algebra is only given by the Carroll-BMS algebra.

\subsection{Electric Carrollian gravity coupled to the electric limit of the Yang-Mills field}

In this case, the theory is invariant under the semi-direct sum of $so(3)$ with the direct sum of the infinite-dimensional sets of generalized Carrollian supertranslations (extended $iso(3)$-Carroll algebra) and color charges.

The asymptotic Killing vector and the non-Abelian gauge parameter that preserve the fall-off of the fields read
\begin{eqnarray}\label{eq:diff-EYM2}
\xi &=&br +T+\mathcal{O}\left(r^{-2}\right)\,,\\
\xi^r &=&W+\mathcal{O}\left(r^{-1}\right)\,,\\
\xi^A &=&Y^A+\frac{1}{r}\left(\xbar D^A W+\frac{2b}{\sqrt{g}}\xbar \pi^{rA}\right)+\mathcal{O}\left(r^{-2}\right)\,,\\
\varepsilon &=&\xbar \varepsilon+\mathcal O(r^{-1})\,,
\end{eqnarray}
where $\xbar D_A \xbar D_B b+\xbar g_{AB}b=0$ and $\xbar D_A Y_B+\xbar D_B Y_A=0$. 

The boundary term of the canonical generator is given by
 \begin{align}
    Q_{\xi,\epsilon}&  = \oint d^2x\Big[2Y^A \Big(\xbar \pi^{(2)r}_A+\frac{1}{2}\xbar A^a_A \xbar \pi^r_a\Big)+2I^A \xbar \pi^r_A+2W \xbar \pi^{rr} -\xbar \epsilon^a \xbar \pi^r_a\Big] \,.
\end{align}
The latter can be re-written as
\begin{equation}
Q_{\xi,\epsilon}=\frac{1}{2}b_{ij}M^{ij}+Q_I+Q_{W}+Q^{\text{YM}}_{\epsilon}\,,
\end{equation}
where
\begin{align}
& M^{ij}=\oint d^2x x^{[i}e^{j]A}\Big(\xbar \pi^{(2)r}_A+\frac{1}{2}\xbar A^a_A \xbar \pi^r_a\Big)\,,\\
&Q_I=\oint d^2 x\,I^A\mathcal{W}_A\,\qquad Q_{W}=\oint d^2x \,W\, \mathcal W\,,\\
&Q^{\text{YM}}_{\epsilon}=\oint d^2x\, \xbar \epsilon^a \mathcal{T}_a\,, 
\end{align}
with $\mathcal W_A=2 \xbar \pi^r_A$, $\mathcal{W}=2\xbar \pi^{rr}$ and $\mathcal{T}_a=-\xbar \pi^r_a$.

The brackets of the charges are then given by
\begin{align}
\{M^{ij},M^{kl}\}&=\frac{1}{2}\Big(\delta^{ik}M^{jl}-\delta^{il}M^{jk}-\delta^{jk}M^{il}+\delta^{jl}M^{ik}\Big)\,,\\
\{M^{ij},\mathcal W(x^A)\}&=-\xbar D_B(x^{[i}e^{j]B} \mathcal W(x^A))\,,\\
\{M^{ij},\cW_A(x^A)\}&=\xbar D_B(x^{[i}e^{j]B} \cW_A(x^A))-x^{[i}\xbar D_Ae^{j]B}\cW_B(x^A)\,,\\
\{M^{ij},\mathcal T_a(x^A)\}&=-\xbar D_B(x^{[i}e^{j]B} \mathcal T_a(x^A))\,,\\
\{\mathcal T_a(x^A_1),\mathcal T_b(x^A_2)\}&=f^c_{\,\,\,ab}\mathcal T_c(x^A_1)\delta^{(2)}(x^A_1-x^A_2)\,.
\end{align}

\subsection{Electric Carrollian gravity coupled to the magnetic limit of the Yang-Mills field}

The color charges in this case vanish because of the strict set of parity conditions on the Yang-Mills field.  The asymptotic symmetry algebra is accordingly only given by the semi-direct sum of $so(3)$ with the infinite-dimensional sets of generalized Carrollian supertranslations. The latter explicitly reads
\begin{align}
\{M^{ij},M^{kl}\}&=\frac{1}{2}\Big(\delta^{ik}M^{jl}-\delta^{il}M^{jk}-\delta^{jk}M^{il}+\delta^{jl}M^{ik}\Big)\,,\\
\{M^{ij},\mathcal W(x^A)\}&=-\xbar D_B(x^{[i}e^{j]B} \mathcal W(x^A))\,,\\
\{M^{ij},\cW_A(x^A)\}&=\xbar D_B(x^{[i}e^{j]B} \cW_A(x^A))-x^{[i}\xbar D_Ae^{j]B}\cW_B(x^A)\,.
\end{align}

\section{Conclusions}
\label{sec:Conclusions}

In this paper, we have analyzed the asymptotic structure of the electric and magnetic Carrollian limits of the Enstein-Yang-Mills theory. Our results can be summarized as follows:

\begin{itemize}
\item We have first constructed BMS-like extensions of the Carroll group on purely algebraic grounds, without reference to the dynamics.  We used the 3+1 description of the symmetry, particularly well suited to rescalings of the energy and the linear momentum  involving different powers of $c$.  The same techniques applied to the more familiar non-relativistic case have  led us to a new BMS-like extension of the Galilean and Bargmann algebras.
\item  We have then considered the Carrollian limit of the pure Einstein theory, where we have shown that the electric contraction allows for a larger set of supertranslations involving three functions on the sphere (with definite parity conditions).  Technically, this is because the condition $\xbar h_{rA} =0$ on the leading order of the mixed radial-angular component of the metric, which reduces the number of symmetry generators in the Einstein case, is not needed any more in the electric contraction.
    \item We studied next the asymptotic symmetries of Carrollian Yang-Mills theories and have shown that contrary to its Minkowskian parent \cite{Tanzi:2020fmt}, the electric Carroll contraction admits at spatial infinity an infinite set of angle-dependent color gauge symmetries fully compatible with Carroll spacetime covariance, including the Carroll boosts. The result extends directly to the combined Einstein-Yang-Mills system.

\end{itemize}

Our results can be extended to higher dimensions by following the methods of \cite{Fuentealba:2021yvo,Fuentealba:2022yqt}, as well as to the coupled electromagnetic-massless scalar field system for which the asymptotic analysis has been performed in   \cite{Tanzi:2021xva}.

Another possible future direction is to explore the possibility of realizing the Galilean-BMS algebra \eqref{eq:CommBPGalileo} as an asymptotic symmetry by imposing suitable boundary conditions and parity conditions in Newton-Cartan gravity \cite{Andringa:2010it}.

It is rather interesting that the electric Carroll limit of the Yang-Mills theory admits the infinite-dimensional color symmetry group found at null infinity (but not at spatial infinity) prior to taking the limit.  This property is in line with the general expectation that the Carroll limit describes well the dynamics at null infinity but definitely deserves further study.

\section*{Acknowledgments}
Discussions with Andr\'es Gomberoff, Joaquim Gomis, Alfredo P\'erez and Ricardo Troncoso are gratefully acknowledged. This work was partially supported by FNRS-Belgium (conventions FRFC PDRT.1025.14 and IISN 4.4503.15), as well as by funds from the Solvay Family. P.S-R. has received funding from the Norwegian Financial Mechanism 2014-2021 via the National Science Centre (NCN) POLS grant 2020/37/K/ST3/03390. JS was supported by a Marina Solvay fellowship.

\bibliography{biblio.bib}

\end{document}